\documentclass[12pt]{article}
\usepackage{amsfonts,amssymb}
\newcommand{\p}[1]{(\ref{#1})}
\newcommand{\nn}{\nonumber}
\newcommand{\ba}{\begin{eqnarray}}
\newcommand{\ea}{\end{eqnarray}}
\newcommand{\be}{\begin{equation}}
\newcommand{\ee}{\end{equation}}

\begin{document}

\newpage
\setcounter{page}{0}
\thispagestyle{empty}
\begin{flushright}
{nlin.SI/0311030}\\
\end{flushright}
\vfill

\begin{center}
{\LARGE {\bf Generalized fermionic discrete}}\\[0.3cm]
{\LARGE {\bf Toda hierarchy}}\\[1cm]

{\large V.V. Gribanov$^{a,1}$, V.G. Kadyshevsky$^{b,2}$ and A.S. Sorin$^{b,3}$}
{}~\\
\quad \\

$^{(a)}${{\em Dzhelepov Laboratory of Nuclear Problems,}}\\
$^{(b)}${{\em Bogoliubov Laboratory of Theoretical Physics,}}\\
{{\em Joint Institute for Nuclear Research,}}\\
{\em 141980 Dubna, Moscow Region, Russia}~\quad\\

{}~

{}~

{\sl Dedicated to the memory of Professor I. Prigogine}

\end{center}

\vfill

\centerline{{\bf Abstract}}
\noindent Bi-Hamiltonian structure and Lax pair formulation with
the spectral parameter of the generalized fermionic
Toda lattice hierarchy as well as its bosonic and fermionic symmetries
for different (including periodic) boundary conditions are described.
Its two reductions ---$N=4$ and $N=2$ supersymmetric Toda lattice
hierarchies--- in different (including canonical) bases are investigated.
Its r-matrix description, monodromy matrix, and spectral curves are discussed.

{}~

{}~

{}~

{}~



\vfill

\begin{flushleft}
{\em ~~~E-mail:}\\
{\em 1) gribanov@thsun1.jinr.ru}\\
{\em 2) kadyshev@jinr.dubna.su}\\
{\em 3) sorin@thsun1.jinr.ru}
\end{flushleft}

\newpage

\section{Introduction}
At present, two different non-trivial supersymmetric extensions
of the two-dimensional (2D) infinite bosonic Toda lattice hierarchy are known.
They are the $N=(2|2)$ \cite{Ik,EH,LS2,LSor,KS1} and $N=(0|2)$ \cite{KS1}
supersymmetric Toda lattice
hierarchies. Actually, besides a different number of supersymmetries they
have different bosonic limits which are decoupled systems of two
infinite bosonic Toda lattice hierarchies and single infinite
bosonic Toda lattice hierarchy, respectively. One-dimensional (1D)
reductions of these hierarchies ---$N=4$ and $N=2$
supersymmetric Toda lattice hierarchies--- were studied in \cite{BS,DGS},
while their finite reductions corresponding to different boundary
conditions (e.g., fixed ends, periodic boundary conditions, etc.) were
investigated in \cite{MAO,Andreev,AuS,LS,dls}. Quite recently, a dispersionless
limit of the $N=(1|1)$ supersymmetric Toda lattice hierarchy was
constructed in \cite{KS2,KS3}.

The present paper continues studies of the
above-mentioned hierarchies and is addressed to yet unsolved problems of
constructing their periodic counterparts, bi-Hamiltonian structure in
different (including canonical) bases, $(2m\times 2m)$-matrix and
$4\times 4$-matrix ($3\times 3$-matrix) Lax pair descriptions with the
spectral parameter, r-matrix approach, and spectral curves.

The structure of this paper is as follows. In section 2.1, starting with the
zero-curvature representation we introduce
the 2D generalized fermionic Toda lattice equations and describe their
two reductions related to the $N=(2|2)$ and $N=(0|2)$
supersymmetric Toda lattice equations. Then, in section 2.2, we construct the
bi-Hamiltonian structure of the 1D generalized fermionic Toda lattice
hierarchy, and its fermionic and bosonic Hamiltonians.

Sections 3 and 4 are devoted to the 1D $N=4$ and $N=2$ supersymmetric
Toda lattice hierarchies, respectively. We construct their bi-Hamiltonian
structure in sections 3.1 and 4.1, fermionic symmetries in section 3.2,
and in sections 3.3 and 4.2, we investigate a transition to the canonical
basis which spoils a number of supersymmetries.

In section 5, we consider periodic supersymmetric Toda lattice hierarchies.
Thus, in section 5.1, we construct the $(2m \times 2m)$-matrix zero-curvature
representation with the spectral
parameter for the periodic 2D generalized fermionic Toda lattice hierarchy.
Then, in section 5.2, we obtain the bi-Hamiltonian structure of its one-dimensional
reduction. In section 5.3, we construct the $(4\times 4)$-matrix Lax pair
representation of this hierarchy, calculate its r-matrix, and analyze monodromy
matrix. We next calculate its spectral curves in section 5.4. In section 5.5,
we give a short summary of the ($3\times 3$)-matrix Lax pair representation and
the r-matrix formalism for the periodic 1D $N=2$ Toda lattice hierarchy, and
calculate spectral curves of the latter. In section 5.6, we discuss periodic Toda lattice
equations in the canonical basis and their fermionic symmetries.


\section{Generalized fermionic Toda lattice hierarchy}

\subsection{2D generalized fermionic Toda lattice equations}
\label{GFTL}
In this subsection we define two-dimensional
generalized fermionic Toda lattice equations
and describe their two different representations which being reduced
relate them with the
$N=(2|2)$ \cite{EH,dls} and $N=(0|2)$ \cite{dls,KS1} supersymmetric Toda
lattice equations.

Our starting point is the following zero-curvature representation:
\be \label{L-cur}
[\partial_1+L^-,\partial_2 -L^+]=0
\ee
 for the infinite matrices
\ba\label{L-mat}
(L^-)_{i,j}=\rho_i \delta_{i,j+1} +d_i \delta_{i,j+2}, && \ \ \ \ \
(L^+)_{i,j}=\delta_{i,j-2}+\gamma_i
\delta_{i,j-1}+c_i \delta_{i,j},
\ea
\ba
 L^-&=&\left(
\begin{array}{cccccccc}
\dots&\dots&&&&&\dots&\dots\\
\dots&0&0&0&0&0&0&\dots\\
&\rho_{j+1}&0&0&0&0&0&\\
&d_{j+2}&\rho_{j+2}&0&0&0&0&\\
&0&d_{j+3}&\rho_{j+3}&0&0&0&\\
&0&0&d_{j+4}&\rho_{j+4}&0&0&\\
\dots&0&0&0&d_{j+5}&\rho_{j+5}&0&\dots\\
\dots&\dots&&&&&\dots&\dots
\end{array}
\right),\nn\\
&&~ \nn\\
 L^+&=&\left(
\begin{array}{cccccccc}
\dots&\dots&&&&&\dots&\dots\\
\dots&c_j&\gamma_j&1&0&0&0&\dots\\
&0&c_{j+1}&\gamma_{j+1}&1&0&0&\\
&0&0&c_{j+2}&\gamma_{j+2}&1&0&\\
&0&0&0&c_{j+3}&\gamma_{j+3}&1&\\
&0&0&0&0&c_{j+4}&\gamma_{j+4}&\\
\dots&0&0&0&0&0&c_{j+5}&\dots\\
\dots&\dots&&&&&\dots&\dots
\end{array}
\right).\nn
\ea
Here, $z_1$ and $z_2$ are the bosonic coordinates
($\partial_{1,2}\equiv \frac{\partial}{\partial z_{1,2}}$);
the matrix entries
$d_j, c_j$ ($\rho_j, \gamma_j$) are the bosonic (fermionic) fields
with Grassmann parity $0$ ($1$)
and length dimensions
$[d_j]=-2$, $[c_j]=-1$, $[\rho_j]=-3/2$ and $[\gamma_j]=-1/2$. The
zero-curvature representation (\ref{L-cur}) leads to the  following
system of evolution equations with respect to the bosonic evolution derivatives
$\partial_{1,2}$:
\ba \label{TODA}
&&\partial_2d_j=d_j(c_j-c_{j-2}),\ \ \
\partial_1c_j=d_{j+2}-d_j+\gamma_j \rho_{j+1}+\gamma_{j-1}\rho_j, \nn  \\
&&\partial_1\gamma_j=\rho_{j+2}-\rho_j,\ \ \
\partial_2\rho_j=\rho_j(c_j-c_{j-1})+d_{j+1} \gamma_j-d_j\gamma_{j-2}.
\label{stoda1}
\ea
Keeping in mind that in the bosonic limit (i.e., when all fermionic fields
are put equal to zero) these equations describe a system of two decoupled
bosonic 2D Toda lattices, we call equations (\ref{stoda1})
the 2D generalized fermionic Toda lattice equations.

Our next goal is to describe fermionic symmetries of the
2D generalized fermionic Toda lattice equations \p{stoda1}.
Before doing so let us first
supply the fields
$(d_j,c_j,\gamma_j,\rho_j)$ with boundary conditions.
In what follows we consider
the boundary conditions of the following four types:
\ba\label{TODA-bound}
I).&&  \lim_{j\to\pm\infty}d_j=0,\ \ \  \
\lim_{j\to\pm\infty}c_j=0,\ \ \  \
\lim_{j\to\pm\infty}\gamma_j=0,\ \ \ \
\lim_{j\to\pm\infty}\rho_j=0;\nn\\
II).&&  \lim_{j\to\pm\infty}d_j=1,\ \ \  \
\lim_{j\to\pm\infty}c_j=0,\ \ \  \
\lim_{j\to\pm\infty}\gamma_j=0,\ \ \ \
\lim_{j\to\pm\infty}\rho_j=0;\nn\\
III).&&  \lim_{j\to\pm\infty}d_{2j+1}=1,\ \ \
 \lim_{j\to\pm\infty}d_{2j}=0,\ \ \
\lim_{j\to\pm\infty}c_j=0,\ \ \
\lim_{j\to\pm\infty}\gamma_j=0,\ \ \
\lim_{j\to\pm\infty}\rho_j=0;\nn\\
IV). &&d_j=d_{j+n}, \ \ \ c_j=c_{j+n}, \ \ \ \gamma_j=\gamma_{j+n}, \ \ \
\rho_j=\rho_{j+n},\ \ \ \ n \in {\mathbb Z}.
\ea
The first three types specify the behavior of the fields at the lattice
points at infinity
while the boundary condition of the fourth type is periodic and corresponds
to the closed 2D generalized fermionic Toda lattice.

For the boundary conditions $I)$ and $II)$ \p{TODA-bound}
the above described equations \p{stoda1} possess the $N=(2|2)$ supersymmetry.
Indeed, in this case there exist four fermionic symmetries
of equations (\ref{TODA})
\ba\label{TODA-susy-12}
&&
\begin{array}{lcl}
D_1^1d_j=  g_{j - 1}   \rho_j + g_j   \rho_{j - 1},&~~~~~~~~~~&
  D_2^1d_j= (-1)^j ( g_{j - 1}   \rho_j - g_j   \rho_{j - 1}),\\
  D_1^1c_j = g_j   \gamma_{j - 1} + g_{j + 1}   \gamma_j,&&
  D_2^1
      c_j   = (-1)^j  ( g_{j + 1}    \gamma_j - g_j   \gamma_{j - 1} ) ,\\
  D_1^1\rho_j = -\partial_1 g_j, &&
  D_2^1 \rho_j   = ( -1 )^j  \partial_1 g_j ,\\
  D_1^1\gamma_j=  g_j - g_{j + 2} ,&&
  D_2^1 \gamma_j   =  ( -1 )^j  ( g_{ j + 2}  - g_j)
\end{array}
\\
\nn\\
\label{TODA-susy-34}
&&
\begin{array}{lcl}
D_3^2 d_j=d_j(\gamma_{j - 1}+ \gamma_{ j - 2}),&~~~~~~~~~~&
  D_4^2
      d_j   =  ( -1 ) ^j d_j    (\gamma_{ j - 1}  - \gamma_{ j - 2 }),\\
D_3^2c_j=
               \partial_2 \sum\limits_{k=- \infty}^{j-1}\gamma_k ,&&
  D_4^2 c_j=
            -\partial_2 \sum\limits_{k=-\infty}^{j - 1}(-1)^k \gamma_k,\\
 D_3^2\rho_j=d_{j + 1}-d_j - \rho_j\gamma_{j-1},&&
      D_4^2 \rho_j=
                     ( -1 )^j  (d_{j + 1}-d_j-\rho_j\gamma_{ j - 1 }),\\
      D_3^2 \gamma_j = c_{ j + 1}  - c_j,&&
      D_4^2 \gamma_j=( -1 ) ^j  (c_{j + 1} -c_j)
 \end{array}
 \ea
where $D_1^1,D_2^1,D_3^2$ and $D_4^2$ are the fermionic evolution derivatives;
$g_j$ denotes the infinite product
\ba
g_j\equiv\prod\limits_{k=0}^\infty
 {\displaystyle\frac{d_{j-2k}}{d_{j-2k-1}}}
\label{composite}
\ea
with the properties $g_jg_{j-1}=d_j$ and
$$  D_1^1g_j =  \rho_j,
\quad  D_2^1 g_j   =  ( -1 ) ^j \rho_j, \quad
  D_3^2g_j = g_j\gamma_{j-1}, \quad
  D_4^2g_j =(-1)^j g_j\gamma_{j-1}, \\
\partial_2 g_j=g_j(c_{j}-c_{j-1}).$$
Now using eqs. (\ref{TODA}) and (\ref{TODA-susy-12})--(\ref{TODA-susy-34})
one can easily check that the
bosonic and fermionic evolution derivatives
satisfy the
algebra of the $N=(2|2)$ supersymmetry
\ba
[\partial_a,\partial_b]=[\partial_a,D_s^b]=0,\ \ \ \ \
\{D_s^1,D_p^1\}=(-1)^s 2 \delta_{s,p}\partial_1,\ \ \ \ \
\{D_s^2,D_p^2\}= -(-1)^s 2\delta_{s,p}\partial_2
\ea
which can be realized via
\ba
 \partial_a =
\frac{\partial}{\partial z_{a}},\ \ \ \ \
D_s^1=
\frac{\partial}{\partial_{\theta_s}}+(-1)^s
\theta_s
\frac{\partial}{\partial z_{1}},\ \ \ \ \
D_p^2=
\frac{\partial}{\partial_{\theta_p}}-(-1)^p
\theta_p
\frac{\partial}{\partial z_{2}}
\ea
where $z_a$ $(a=1,2)$
and $\theta_{s},\theta_{p}$ $(s=1,2;\,$ $p=3,4)$ are the bosonic
and fermionic evolution
times of the $N=(2|2)$ superspace, respectively.

Looking at equations (\ref{TODA-susy-12})--(\ref{TODA-susy-34})
one can see that they are not consistent with the boundary conditions
$III)$ \p{TODA-bound}.
Thus, it is impossible to simultaneously
satisfy the boundary conditions for the fields $g_j$
entering into eqs. (\ref{TODA-susy-12})
\ba \label{bc}
 \lim_{j\to\pm\infty}d_{2j}=
 \lim_{j\to\pm\infty}g_{2j}g_{2j-1}=0,\ \ \ \ \
 \lim_{j\to\pm\infty}d_{2j+1}=
 \lim_{j\to\pm\infty}g_{2j+1}g_{2j}=1,
\ea
while eqs. (\ref{TODA-susy-34}) contain a contradiction at infinity
in the equation for the field $\rho_j.$
Thus, one can conclude
that the boundary conditions strictly restrict the  symmetries of eqs.(\ref{TODA}).
The periodic boundary conditions will be considered in section 5.

Now we present other two related representations of
the 2D generalized fermionic Toda lattice equations \p{stoda1}
which will be useful in what follows.

The first representation can
easily be derived if one introduces a new basis
$\{g_j,c_j,\gamma^+_j,\gamma^-_j\}$
in the space of the fields
$\{d_j,c_j,\gamma_j,\rho_j\}$
\ba\label{g-basis}
d_j=g_jg_{j-1},\ \ \ \ \rho_j=g_j\gamma^-_j,\ \ \ \
\gamma_j=\gamma_{j+1}^+
\ea
and eliminate the fields $c_j$ from eq. (\ref{TODA}) in order to get the
conventional form of the 2D $N=(2|2)$ supersymmetric Toda lattice equations
\cite{dls}
\begin{eqnarray}\label{g-toda}
\partial_1\partial_2 \mbox{\rm ln}g_j&=&
g_{j+1}g_{j+2}-g_j(g_{j+1}+g_{j-1})+
g_{j-1}g_{j-2}+g_{j+1}\gamma^+_{j+1}\gamma^-_{j+1}-
g_{j-1}\gamma_{j-1}^+\gamma_{j-1}^-,\nn \\
\partial_1\gamma_j^+&=&g_{j+1}\gamma_{j+1}^--g_{j-1}\gamma_{j-1}^-, \ \ \ \ \ \
\partial_2\gamma_j^-=g_{j+1}\gamma_{j+1}^+-g_{j-1}\gamma_{j-1}^+.
\end{eqnarray}
together with their fermionic $N=(2|2)$ symmetries
\ba\label{susy-g-toda}
\begin{array}{lcl}
D_1^1g_j =g_j \gamma_j^-,&~~~~~~~~~~~~&
D_2^1g_j = (-1)^j g_j \gamma_j^-,\\
D_1^1\gamma_j^- =-\partial_1  \mbox{\rm ln}g_j,&&
D_2^1\gamma_j^- =(-1)^j \partial_1  \mbox{\rm ln}g_j,\\
 D_1^1\gamma_j^+ =g_{j-1}- g_{j+1},&&
 D_2^1\gamma_j^+ =(-1)^j (g_{j-1}- g_{j+1}),
\end{array}
\nn\\
\begin{array}{lcl}
D_3^2g_j =g_j \gamma_j^+,&~~~~~~~~~~~~&
D_4^2g_j =(-1)^j g_j \gamma_j^+, \\
D_3^2\gamma_j^- =g_{j+1}- g_{j-1},&&
 D_4^2\gamma_j^- =(-1)^j ( g_{j+1}- g_{j-1}),\\
D_3^2\gamma_j^+ =\partial_2  \mbox{\rm ln} g_j,&&
D_4^2\gamma_j^+ =-(-1)^j \partial_2  \mbox{\rm ln} g_j.
\end{array}
\ea

In order to derive the second representation, let us
introduce a new notation for the fields at
odd and even values of the lattice coordinate $j$
\begin{eqnarray}\label{ab-basis}
&& a_j\equiv c_{2j+1},\ \ \ b_j\equiv d_{2j+1},\ \ \ \alpha_j\equiv
\gamma_{2j-1}, \ \ \ \beta_j\equiv \rho_{2j+1}, \nn \\
&&\bar a_j\equiv c_{2j}, \ \ \ \bar b_j\equiv d_{2j},\ \ \
\bar \alpha_j\equiv -\gamma_{2j},\ \ \ \bar \beta_j\equiv \rho_{2j}
\end{eqnarray}
and rewrite eqs. (\ref{TODA}), (\ref{TODA-susy-12}--\ref{TODA-susy-34})
in the following form:
\begin{eqnarray}  \label{ab-toda}
&&\partial_2b_j=b_j(a_j-a_{j-1}),\ \ \
\partial_1a_j=
b_{j+1}-b_j+\beta_j\bar\alpha_j+ \alpha_{j+1}\bar\beta_{j+1},\nn \\
&&\partial_2\bar b_j=\bar b_j(\bar a_{j}-\bar a_{j-1}),\ \ \
\partial_1\bar a_j=
\bar b_{j+1}-\bar b_{j}+\beta_j\bar\alpha_j+ \alpha_{j}\bar\beta_{j},\nn\\
&&\partial_1\alpha_j=\beta_j-\beta_{j-1},\ \ \
 \partial_2\beta_j=(a_j-\bar a_j)\beta_j -b_j \alpha_j+\bar
b_{j+1}\alpha_{j+1},\nn\\
&&\partial_1\bar\alpha_j=\bar\beta_j-\bar\beta_{j+1},\ \ \
\partial_2\bar\beta_j=(\bar a_{j}- a_{j-1})\bar
\beta_j-b_j\bar\alpha_j+\bar b_{j}\alpha_{j-1},
 \end{eqnarray}
\ba\label{toda-ab-susy-12}
&&
\begin{array}{lcl}
D_1^1b_j=e_j\bar\beta_j+\bar e_j\beta_j,&~~~~~~~~&
D_2^1b_j=-e_j\bar\beta_j-\bar e_j\beta_j,\\
D_1^1\bar b_j=e_{j-1}\bar\beta_j+\bar e_j\beta_{j-1},&~~~~~~~~&
D_2^1\bar b_j=e_{j-1}\bar\beta_j-\bar e_j\beta_{j-1},\\
D_1^1a_j=\bar e_{j+1}\alpha_{j+1}-e_{j}\bar\alpha_{j},&&
D_2^1a_j=-\bar e_{j+1}\alpha_{j+1}-e_{j}\bar\alpha_{j},\\
D_1^1\bar a_j=\bar e_{j}\alpha_{j}-e_{j}\bar\alpha_{j},&&
D_2^1\bar a_j=-\bar e_{j}\alpha_{j}-e_{j}\bar\alpha_{j},\\

D_1^1 \beta_j=-\partial_1 \bar e_j,&&
D_2^1 \beta_j=-\partial_1 \bar e_j,\\
D_1^1 \bar\beta_j=-\partial_1 \bar e_j,&&
D_2^1 \bar\beta_j=\partial_1 \bar e_j,\\
D_1^1\alpha_j=e_{j-1}-e_{j},&&
D_2^1\alpha_j=e_{j-1}-e_{j},\\
D_1^1\bar\alpha_j=\bar e_{j+1}-\bar e_{j},&&
D_2^1\bar \alpha_j=\bar e_{j}-\bar e_{j+1},
\end{array}
\\
&&
\nn\\
&& \label{toda-ab-susy-34}
\begin{array}{lcl}
D_3^2b_j=b_j(\alpha_j-\bar \alpha_j),&~~~~~~~~&
D_4^2b_j=b_j(\alpha_j+\bar \alpha_j),\\
D_3^2\bar b_j=\bar b_j(\alpha_j-\bar \alpha_{j-1}),&&
D_4^2\bar b_j=\bar b_j(\alpha_j+\bar \alpha_{j-1}),\\
D_3^2 a_j=\partial_2 \sum\limits_{k=-\infty}^{j}
(\alpha_k-\bar \alpha_k),&~~~~~~~~&
D_4^2a_j=\partial_2 \sum\limits_{k=-\infty}^{j}
(\alpha_k+\bar \alpha_k),\\
D_3^2 \bar a_j=\partial_2 \sum\limits_{k=-\infty}^{j}
(\alpha_k-\bar \alpha_{k-1}),&~~~~~~~~&
D_4^2\bar a_j=\partial_2 \sum\limits_{k=-\infty}^{j}
(\alpha_k+\bar \alpha_{k-1}),\\
D_3^2\beta_j=\bar b_j-b_j-\bar b_i+\beta_j\bar\alpha_j,&&
D_4^21\beta_j=b_j-\bar b_{j+1}-\beta_j\bar\alpha_j,\\
D_3^2\bar\beta_j=b_j-\bar b_j-\bar\beta_j\alpha_j,&&
D_4^2\bar\beta_j=b_j-\bar b_j-\bar\beta_j\alpha_j,\\
D_3^2\alpha_j=\bar a_j-a_{j-1},&&
D_4^2\alpha_j=a_{j-1}-\bar a_j,\\
D_3^2\bar\alpha_j=\bar a_j-a_{j},&&
D_4^2\bar\alpha_j=\bar a_j-a_{j}
\end{array}
\ea
where  $e_j,\bar e_j$ are the composite fields
\ba\label{e-repr}
e_j\equiv g_{2j+1}\equiv\prod\limits_{k=0}^\infty
 {\displaystyle\frac{b_{j-k}}{\bar b_{j-k}}},
\ \ \ \ \ \ \
\bar e_j\equiv g_{2j}\equiv\prod\limits_{k=0}^\infty
 {\displaystyle\frac{\bar b_{j-k}}{ b_{j-k-1}}}
\ea
which obey the equations
\ba
&&\partial_2 e_j=e_j(a_j-\bar a_j),\ \ \
\partial_2 \bar e_j=\bar e_j(\bar a_j- a_{j-1}),\nn\\
&&D_1^1e_j=\beta_j,\ \ \
D_2^1e_j=-\beta_j,\ \ \
D_3^2e_j=e_j \alpha_j,\ \ \
D_4^2e_j=-e_j \alpha_j,\nn\\
&&D_1^1\bar e_j=\bar\beta_j,\ \ \
D_2^1\bar e_j=\bar \beta_j,\ \ \
D_3^2\bar e_j=-\bar e_j \bar\alpha_j,\ \ \
D_4^2\bar e_j=-\bar e_j \bar \alpha_j.
\ea
The reduction
\ba
{\bar b}_j =0
\label{reds1}
\ea
of eqs. \p{ab-toda} leads to
the 2D $N=(0|2)$ supersymmetric Toda lattice equations
\cite{dls,KS1}. One can easily see that fermionic symmetries
\p{toda-ab-susy-12} are not consistent with this reduction, while
fermionic symmetries \p{toda-ab-susy-34} are consistent and
form the algebra of the $N=(0|2)$ supersymmetry.

\subsection{Bi-Hamiltonian structure of the 1D generalized fermionic Toda
lattice hierarchy}
\label{1d-GFTL}
Our further purpose is to
construct a bi-Hamiltonian structure of the generalized fermionic Toda lattice
equations (\ref{TODA})
(and, consequently, originating from them eqs. (\ref{g-toda}) and (\ref{ab-toda}))
in one-dimensio\-nal space when all the fields
depend on only one bosonic coordinate $z=z_1+z_2$.
This task was solved in \cite{BS}
for the 1D $N=2$ Toda lattice hierarchy obtained by reduction \p{reds1}
of the 1D generalized fermionic Toda lattice hierarchy. Here we solve
this task for the original 1D generalized fermionic Toda lattice hierarchy.

At the reduction to one-dimensional space,
\ba
\partial_1=\partial_2\equiv\partial,
\label{red-one}
\ea
the zero-curvature representation (\ref{L-cur}) can
 identically be rewritten in the form of the Lax-pair
representation
\begin{equation}\label{L-pair}
\partial L=[L,L^-], \ \ \ \ L\equiv L^++L^-,
\end{equation}
\vspace*{-.8cm}
\ba
 L&=&\left(
\begin{array}{cccccccc}
\dots&\dots&&&&&\dots&\dots\\
\dots&c_j&\gamma_j&1&0&0&0&\dots\\
&\rho_{j+1}&c_{j+1}&\gamma_{j+1}&1&0&0&\\
&d_{j+2}&\rho_{j+2}&c_{j+2}&\gamma_{j+2}&1&0&\\
&0&d_{j+3}&\rho_{j+3}&c_{j+3}&\gamma_{j+3}&1&\\
&0&0&d_{j+4}&\rho_{j+4}&c_{j+4}&\gamma_{j+4}&\\
\dots&0&0&0&d_{j+5}&\rho_{j+5}&c_{j+5}&\dots\\
\dots&\dots&&&&&\dots&\dots
\end{array}
\right).\nn
\ea
Using the Lax pair representation (\ref{L-pair}),
it is easy to derive the general expression for bosonic Hamiltonians
which are in involution via the standard formula
\begin{equation}\label{str}
H_k=\frac1k str L^k\equiv\frac1k \sum_{p=1}^{\infty} (-1)^p (L^k)_{pp}.
\end{equation}
The first two of them have the following explicit form:
\ba\label{HAM-bos}
 &&H_1 = \sum_{i =- \infty }^{\infty} (-1)^i c_i,\ \ \ \ \ \
H_2 = \sum_{i =-\infty}^{\infty} (-1)^i (  \frac12 c_i^2 +d_i+
 \rho_i \gamma_{i- 1}).
\ea

A bi-Hamiltonian system of evolution equations can be represented in the
following general form:
\ba \label{flows}
\frac{\partial}{\partial t_{H_k}}~ q_i&=&
\{{ H}_{k+1},q_i\}_1=\{{ H}_k,q_i\}_2
\ea
where $t_{H_k}$ are the evolution times, $q_j$ denotes any field from the set
$q_i=\{d_i,c_i,\rho_i,\gamma_i\}$ and the brackets $\{,\}_{1(2)}$
are appropriate Poisson brackets corresponding to the first (second)
Hamiltonian structure. Using eqs. \p{flows} and the 2D generalized fermionic
Toda lattice equations (\ref{TODA}) at the reduction to one-dimensional space
\p{red-one} -- the 1D generalized fermionic
Toda lattice equations
\ba \label{TODA-1d}
&&\partial d_i=d_i(c_i-c_{i-2}),\ \ \
\partial c_i=d_{i+2}-d_i+\gamma_i \rho_{i+1}+\gamma_{i-1}\rho_i, \nn  \\
&&\partial \gamma_i=\rho_{i+2}-\rho_i,\ \ \
\partial \rho_i=\rho_i (c_i-c_{i-1})+d_{i+1}
\gamma_i-d_i\gamma_{i-2}
\ea
as well as Hamiltonians \p{HAM-bos},
we have found the first
two Hamiltonian structures  of the hierarchy. As the result, we have
the following explicit expressions:
 \ba\label{1pb-TODA}
\{d_i, c_j\}_1& =&  ( -1 ) ^j d_i( \delta_{i, j+2}  - \delta_{i, j }),\nn \\
\{c_i,\rho_j\}_1& =&  ( -1 ) ^j \rho_j
        (\delta_{i, j - 1} +  \delta_{i, j}),\nn \\
  \{\rho_i,
      \rho_j\}_1 &=&      ( -1 ) ^j (d_i \delta_{i,
              j + 1}\  - d_j \delta_{i,j - 1})   ,\nn\\
  \{\gamma_i, \gamma_j\}_1& =&
     ( - 1 ) ^j ( \delta_{i, j + 1}- \delta_{i,j - 1})
\ea
for the first and
\ba\label{2pb-TODA}
  \{d_i,d_j\}_2&=&(-1)^j d_i d_j(\delta_{i,j+2}-\delta_{i,j-2}),\nn \\
  \{d_i, c_j\}_2&=&( -1 )^j d_i c_j ( \delta_{i, j+2}- \delta_{i,j}),\nn \\
  \{c_i,c_j\}_2&=&(-1)^j (d_i \delta_{i, j + 2} -d_j \delta_{i, j - 2} -
            \gamma_j  \rho_i \delta_{i, j+1}-
                 \gamma_i \rho_j \delta_{i, j - 1}),\nn\\
 \{d_i,\rho_j\}_2&=&(-1)^j  d_i \rho_j (\delta_{i,j+2}+\delta_{i,j-1}),\nn\\
  \{d_i,\gamma_j\}_2&=&(-1)^j d_i
        \gamma_j  ( \delta_{i, j + 2} +\delta_{i, j + 1}),\nn\\
  \{c_i,\rho_j\}_2&=&(-1)^j (c_i  \rho_j (\delta_{i,j}+\delta_{i,j-1})-
     d_j  \gamma_i \delta_{i, j - 2} - d_i  \gamma_j\ \delta_{i, j + 1}),\nn\\
  \{c_i,\gamma_j\}_2&=&(-1)^j (\rho_i \delta_{i, j + 2} + \rho_j
        \delta_{i, j - 1}),\nn\\
 \{\rho_i,\gamma_j\}_2&=&(-1)^j  (\rho_i  \gamma_j \delta_{i, j + 1}+
              d_i \delta_{i,j + 3} - d_j \delta_{i, j - 1}),\nn\\
\{\rho_i,
      \rho_j\}_2&=&(-1)^j ((\rho_i  \rho_j - d_j  c_i)\delta_{i,j - 1}+
     (\rho_i  \rho_j + d_i  c_j) \ \delta_{i, j + 1}),\nn\\
 \{\gamma_i,\gamma_j\}_2&=&( -1 ) ^j(  c_i
          \delta_{i, j+1}- c_j\delta_{i,j - 1})
\ea
for the second Hamiltonian structures, where only nonzero brackets are
written down.

Note that the first
$\{,\}_1$ (\ref{1pb-TODA}) and the second $\{,\}_2$ (\ref{2pb-TODA})
Hamiltonian structures   are obviously compatible:
the deformation of the fields $c_j\to c_j+\nu$, where $\nu$ is an
arbitrary constant, transforms $\{,\}_2$ into the Hamiltonian
structure which is their sum
$$\{,\}_2\ \to \ \{,\}_2+\nu\ \{,\}_1.$$
Thus, one concludes that the corresponding recursion operator
$$R=\{,\}_2 \{,\}^{-1}_1$$
is hereditary like
the operator obtained from the compatible pair of Hamiltonian
structures.

We have checked that the one-dimensional reduction \p{red-one}
of the fermionic symmetries
\p{TODA-susy-12}--\p{TODA-susy-34}
\ba\label{TODA-1d-susy}
&&
\begin{array}{lcl}
D_1d_i=  g_{i - 1}   \rho_i + g_i   \rho_{i - 1},&~~~~~~~~~~&
  D_2d_i= (-1)^i ( g_{i - 1}   \rho_i - g_i   \rho_{i - 1}),\\
  D_1c_i = g_i   \gamma_{i - 1} + g_{i + 1}   \gamma_i,&&
  D_2
      c_i   = (-1)^i  ( g_{i + 1}    \gamma_i - g_i   \gamma_{i - 1} ) ,\\
  D_1\rho_i =  g_i (c_{i-1}-c_i), &&
  D_2 \rho_i   = ( -1 )^i g_i(c_i-c_{i-1})  ,\\
  D_1\gamma_i=  g_i - g_{i + 2} ,&&
  D_2 \gamma_i   =  ( -1 )^i  ( g_{ i + 2}  - g_i)
\end{array}
\nn\\
&&\begin{array}{lcl}
D_3 d_i=d_i(\gamma_{i - 1}+ \gamma_{ i - 2}),&~~~~~~~~~~&
  D_4
      d_i   =  ( -1 ) ^i d_i    (\gamma_{ i - 1}  - \gamma_{ i - 2 }),\\
D_3c_i=
                          \rho_{i+1}+\rho_i ,&&
  D_4 c_i=
                               (-1)^i ( \rho_{i+1}-\rho_i),\\
 D_3\rho_i=d_{i + 1}-d_i - \rho_i\gamma_{i-1},&&
      D_4 \rho_i=
                     ( -1 )^i  (d_{i + 1}-d_i-\rho_i\gamma_{ i - 1 }),\\
      D_3 \gamma_i = c_{ i + 1}  - c_i,&&
      D_4 \gamma_i=( -1 ) ^i  (c_{i + 1} -c_i)
 \end{array}
 \ea
and the equations for the composite fields $g_i$ \p{composite}
\ba\label{g-eq}
\partial g_j=g_j(c_{j}-c_{j-1}),\ \
  D_1g_j =  \rho_j,\ \
  D_2 g_j   =  ( -1 ) ^j \rho_j,\  \
  D_3g_j = g_j\gamma_{j-1}, \ \
  D_4g_j =(-1)^j g_j\gamma_{j-1}~
\ea
can also be represented in
a bi-Hamiltonian form with fermionic Hamiltonians $S_{s,k}$ and Hamiltonian
structures (\ref{1pb-TODA}) and (\ref{2pb-TODA})
\ba \label{flows-fer}
D_{t_{S_{s,k}}} q_i&=&
\{S_{s,k+1},q_i\}_1=\{S_{s,k},q_i\}_2
\ea
where $D_{t_{S_{s,k}}}$ are the fermionic evolution derivatives.
In  section
\ref{FerH}
we show how fermionic Hamiltonians can be
derived in an algorithmic way, but now
let us only mention that there are four infinite towers of
fermionic Hamiltonians $S_{s,k}$ $(s=1,2,3,4; ~k\in\mathbb N$)
and present without any comments only
explicit expressions for the  first  few of them
\ba\label{HAM-fer}
&&  S_{1,1} = \sum_{i = -\infty}^{\infty} (-1)^i \rho_i
g_i^{-1},\ \ \
S_{1,2} =- \sum_{i = -\infty}^{\infty} \left( (-1)^i \ g_i
  \gamma_{i - 1} + \rho_i  g_i^{-1}   \sum_{j = -\infty}^{i-1}(-1)^j
        c_j\right),\nn\\
&& S_{2,1} = \sum_{i = -\infty}^{\infty}
\rho_i  g_i^{-1},\ \ \
S_{2,2} = \sum_{i = -\infty}^{\infty} \left(  \ g_i
  \gamma_{i - 1} - (-1)^i  \rho_i  g_i^{-1}   \sum_{j = -\infty}^{i-1}(-1)^j
      c_j\right),\nn\\
&& S_{3,1} = -\sum_{i = -\infty}^{\infty} (-1)^i
\gamma_i,\ \ \
S_{3,2} = -\sum_{i=-\infty}^{\infty}\left (
    (-1)^i\rho_i+\gamma_{i-1}  \sum_{j=-\infty}^{i-1} (-1)^j\ c_j
    \right ),\nn\\
&& S_{4,1} = \sum_{i = -\infty}^{\infty}  \gamma_i,\ \
\ S_{4,2} = \sum_{i=-\infty}^{\infty}\left( \rho_i-(-1)^i\gamma_{i-1}
     \sum_{j=-\infty}^{i-1} (-1)^j\ c_j \right).
 \ea
For completeness we also present the nonzero Poisson brackets
of the composite field $g_i$ \p{composite} with other fields of the hierarchy
which are useful when producing fermionic Hamiltonian flows
\ba\label{g-PB}
 \{g_i,c_j\}_1&=&(-1)^j  g_i (\delta_{i,j+1}-\delta_{i,j}),\nn\\
\{g_i,\gamma_j\}_2&=&(-1)^j g_i \gamma_j \delta_{i,j+1},\nn\\
\{g_i,c_j\}_2&=&(-1)^j g_i c_j(
\delta_{i,j+1}-\delta_{i,j}),\nn\\
\{g_i,\rho_j\}_2&=&(-1)^j g_i \rho_j
                 (\delta_{i,j+1}-\delta_{i,j}+\delta_{i,j-1}),\nn\\
\{g_i,d_j\}_2&=&(-1)^j g_i d_j (
\delta_{i,j+1}-\delta_{i,j}+\delta_{i,j-1}-\delta_{i,j-2}),\nn\\
\{g_i,g_j\}_2&=&(-1)^j g_i g_j( \delta_{i,j+1}+\delta_{i,j-1}).
\ea

Now we have all necessary ingredients to derive Hamiltonian
flows of the 1D generalized Toda lattice hierarchy. Let us end
this section with a few remarks.

First, the Hamiltonians $H_1$ \p{HAM-bos} and $S_{s,1}$ \p{HAM-fer}
give trivial flows via
the first Hamiltonian structure (\ref{1pb-TODA}) because
they belong to the center of the
algebra (\ref{1pb-TODA})
\ba
\{H_1,q_j\}_1=\{S_{s,1},q_j\}_1=0.
\ea

Second, while the densities
corresponding to the fermionic Hamiltonians $S_{p,k}$ (\ref{HAM-fer})
have a nonlocal character with respect to the lattice indices, the fermionic
flows (\ref{TODA-1d-susy}) have no nonlocal terms.

Finally, the algebras of the first and second Hamiltonian structures
(\ref{1pb-TODA})--(\ref{2pb-TODA}) together with eqs. (\ref{g-PB})
possess a discrete inner automorphism $f$ which transforms nontrivially
only fermionic fields
\ba\label{toda-auto}
\gamma_j\ \stackrel{f}{\longmapsto}\ (-1)^j\rho_{j+1}g_{j+1}^{-1}, \ \ \ \
\rho_j\ \stackrel{f}{\longmapsto}\ (-1)^j\gamma_{j-1}g_{j}.
\ea
Using eqs. (\ref{g-eq}) one can easily check that the automorphism
$f$ transforms eqs. (\ref{TODA-1d}), (\ref{TODA-1d-susy})
and Hamiltonians (\ref{HAM-bos}), (\ref{HAM-fer}) according to
the following rule:
\ba
&&\{\partial,D_1,D_2,D_3,D_4\}\ \stackrel{f}{\longmapsto}
\{\partial,D_4,D_3,-D_2,-D_1\},\nn\\
&&\{H_k,S_{1,k},S_{2,k},S_{3,k},S_{4,k}\}\  \stackrel{f}{\longmapsto}
\{H_k,-S_{4,k},-S_{3,k},-S_{2,k},-S_{1,k}\}.
\ea

\section{Reduction: 1D N=4 supersymmetric Toda lattice hierarchy}
\setcounter{equation}{0}
\subsection{Bi-Hamiltonian structure of the 1D N=4 Toda lattice hierarchy}
\label{N4-TL}
In this section we consider the bi-Hamiltonian formulation of the
 one-dimensional reduction \p{red-one} of the
2D $N=(2|2)$ supersymmetric Toda lattice equations (\ref{g-toda})
and their fermionic symmetries \p{susy-g-toda}.
Starting with Hamiltonians (\ref{HAM-bos}), (\ref{HAM-fer}) and
Hamiltonian structures (\ref{1pb-TODA})--(\ref{2pb-TODA})
as well as using relations (\ref{g-basis})
it is easy to represent eqs. (\ref{g-toda})
in one-dimensional space as a bi-Hamiltonian system
of first order evolution equations.

Thus, we obtain the following bosonic and fermionic Hamiltonians:
  \ba \label{g-ham}
  && H_1^{N=4} = \sum_{ i = -\infty} ^{\infty}  ( -1 ) ^i c_i,\ \ \
  H_2^{N=4} =\sum_{ i = -\infty} ^{\infty} (-1)^i \left ( \frac12  c_i^2+ g_i g_{i- 1}  +
g_i \gamma^-_i \gamma^+_i\right),\nn\\
&&
  S_{1,1}^{N=4} = \sum_{i =
  -\infty}^{\infty}(-1)^i \gamma^-_i,\ \ \
  S_{1,2}^{N=4} =- \sum_{i = -\infty} ^{\infty}\left(  (-1)^ig_i
  \gamma^+_i + \gamma^-_i \sum_{ k = -\infty} ^{i-1}(
      -1 )^k  c_k \right),\nn\\
  &&
  S_{2,1}^{N=4} = \sum_{i = -\infty} ^{\infty}
  \gamma^-_i,\ \ \
  S_{2,2}^{N=4} = \sum_{i = -\infty} ^{\infty}\left(
  g_i\gamma^+_i - ( -1 )^i \gamma^-_i  \sum_{ k = -\infty} ^{i-1}( -1 ) ^k
   c_k\right),\nn\\
  &&
  S_{3,1}^{N=4} = \sum_{i=-\infty}^{\infty} (-1)^i \gamma^+_i ,\ \ \
  S_{3,2}^{N=4}= -\sum_{i=-\infty}^{\infty}\left((-1)^i g_i \gamma^-_i +
  \gamma^+_i     \sum_{k=-\infty}^{i-1}( -1 )^k c_k\right),\nn\\
&&
S_{4,1}^{N=4} = \sum_{ i = -\infty} ^{\infty}  \gamma^+_i,\ \ \
S_{4,2}^{N=4} = \sum_{ i = -\infty} ^{\infty}\left( g_i\gamma^-_i
   -(-1)^i\gamma^+_i\sum_{k=-\infty}^{i-1}( -1 ) ^k c_k\right),
\ea
and the first
\ba \label{g-1PB}
  \{\gamma^\pm_i,
      \gamma^\pm_j\}_1&=&\pm  ( -1 ) ^j  (
        \delta_{i, j - 1} - \delta_{i, j + 1}),\nn\\
  \{g_i, c_j\}_1&=&  ( -1 ) ^j g_i  (
        \delta_{i, j + 1} -
          \delta_{i, j})
\ea
and the second
\ba\label{g-2PB}
  \{g_i,
      g_j\}_2 &=&  ( -1 ) ^j g_i
        g_j  (\delta_{i, j + 1} + \delta_{i, j - 1}),\nn\\
  \{g_i, \gamma^\pm_j\}_2&=&
            - ( -1 ) ^j \ \ g_i   \gamma^\pm_j\ \delta_{i, j},\nn\\
 \{\gamma^\pm_i,
      \gamma^\pm_j\}_2&=&\pm  ( -1 )^j(c_i\ \delta_{i, j - 1} -
        c_j \delta_{i, j + 1}), \nn\\
 \{\gamma^-_i,
      \gamma^+_j\}_2&=&  ( -1 ) ^
        j\  (g_ {i + 1}  \delta_{i, j - 2} -
          g_{ i - 1}  \delta_{i, j + 2}), \nn\\
   \{c_i, c_j\}_2&=&  ( -1 ) ^j  (\
        g_i   g_{ i - 1} \delta_{i, j + 2} -  g_j   g_{ j - 1 }
            \delta_{i, j - 2} -
                g_i \gamma^+_i \gamma^-_i \delta_{i,
              j + 1} - g_j \gamma^+_j   \gamma^-_j  \delta_{i,
              j - 1}), \nn\\
  \{g_i,
      c_j\}_2&=&  ( -1 ) ^j\ g_i   c_j  (
        \delta_{i, j + 1} - \ \delta_{i, j}), \nn\\
  \{c_i,
      \gamma^\pm_j\}_2&=&  - ( -1 ) ^
          j  ( g_i   \gamma^\mp_i \delta_{i, j + 1} +  g_{ j - 1   }
              \gamma^\mp_{j - 1} \delta_{i, j - 2})
\ea
Hamiltonian structures,
where only nonzero brackets are presented.
For the first nontrivial bosonic and fermionic  flows one obtains
in a standard way, using eqs. (\ref{flows}), \p{flows-fer} and
\p{g-ham}--\p{g-2PB},
\begin{eqnarray}\label{g-toda-1d-1order}
&&\partial g_i=g_i(c_i-c_{i-1}),\ \ \ \ \ \
\partial c_i=
-g_{i-1}g_{i}+g_{i+1}g_{i+2}
+g_{i+1}\gamma^+_{i+1}\gamma^-_{i+1}+
g_{i}\gamma_{i}^+\gamma_{i}^-,\nn \\
&&\partial\gamma_i^+=g_{i+1}\gamma_{i+1}^--g_{i-1}\gamma_{i-1}^-, \ \ \ \ \ \
\partial\gamma_i^-=g_{i+1}\gamma_{i+1}^+-g_{i-1}\gamma_{i-1}^+,
\end{eqnarray}
\ba\label{g-toda-1d-1order-susy}
&&
\begin{array}{lcl}
D_1g_i =g_i \gamma_i^-,&~~~~~~~~~~~~&
D_2g_i = (-1)^i g_i \gamma_i^-,\\
D_1\gamma_i^- =c_{i-1}-c_i,&&
D_2\gamma_i^- =(-1)^i (c_{i}-c_{i-1}),\\
 D_1\gamma_i^+ =g_{i-1}- g_{i+1},&&
 D_2\gamma_i^+ =(-1)^i (g_{i-1}- g_{i+1}),\\
 D_1c_i =g_{i+1}\gamma_{i+1}^++ g_{i}\gamma^+_i,&&
 D_2c_i =(-1)^i (g_{i+1}\gamma_{i+1}^+- g_{i}\gamma^+_i),
\end{array}
\nn\\
&&
\nn\\
&&
\begin{array}{lcl}
D_3g_i =g_i \gamma_i^+,&~~~~~~~~~~~~&
D_4g_i =(-1)^i g_i \gamma_i^+, \\
D_3\gamma_i^- =g_{i+1}- g_{i-1},&&
 D_4\gamma_i^- =(-1)^i ( g_{i+1}- g_{i-1}),\\
D_3\gamma_i^+ =c_i-c_{i-1},&&
D_4\gamma_i^+ =(-1)^i (c_{i-1}-c_i),\\
 D_3c_i =g_{i+1}\gamma_{i+1}^-+ g_{i}\gamma^-_i,&&
 D_4c_i =(-1)^i( g_{i+1}\gamma_{i+1}^- -g_{i}\gamma^-_i).
\end{array}
\ea
For the system (\ref{g-toda-1d-1order}) we consider the
boundary conditions at infinity of the following two types
\ba\label{22-bc}
Ia).\ \lim_{j\to\pm\infty}g_j=0,\ \ \ \lim_{j\to\pm\infty}c_j=0,\ \ \ \
 \lim_{j\to\pm\infty}\gamma^\pm_j=0;\nn\\
IIa).\ \lim_{j\to\pm\infty}g_j=1,\ \ \ \lim_{j\to\pm\infty}c_j=0,\ \ \
 \lim_{j\to\pm\infty}\gamma^\pm_j=0
\ea
which are the consequences of the
boundary conditions $I)$ and $II)$ \p{TODA-bound}.
Flows (\ref{g-toda-1d-1order})--(\ref{g-toda-1d-1order-susy}) are compatible
with these boundary conditions and form
the $N=4$ supersymmetry algebra.

The fields $c_i$ can be dropped out of the system (\ref{g-toda-1d-1order}) and
finally eqs. (\ref{g-toda-1d-1order}) take the  form of
eqs. (\ref{g-toda}) in one-dimensional space \p{red-one}
\begin{eqnarray}\label{g-toda-1d}
\partial^2 \mbox{\rm ln}g_i&=&
g_{i+1}g_{i+2}-g_i(g_{i+1}+g_{i-1})+
g_{i-1}g_{i-2}+g_{i+1}\gamma^+_{i+1}\gamma^-_{i+1}-
g_{i-1}\gamma_{i-1}^+\gamma_{i-1}^-,\nn \\
&&\partial\gamma_i^+=g_{i+1}\gamma_{i+1}^--g_{i-1}\gamma_{i-1}^-, \ \ \ \ \ \
\partial\gamma_i^-=g_{i+1}\gamma_{i+1}^+-g_{i-1}\gamma_{i-1}^+
\end{eqnarray}
with the $N=4$ supersymmetry transformations
\be
\begin{array}{lcl}
D_1g_i =g_i \gamma_i^-,&~~~~~~~~~~~~&
D_2g_i = (-1)^i g_i \gamma_i^-,\\
D_1\gamma_i^- =-\partial{\rm ln}g_i,&&
D_2\gamma_i^- =(-1)^i \partial{\rm ln}g_i,\\
 D_1\gamma_i^+ =g_{i-1}- g_{i+1},&&
 D_2\gamma_i^+ =(-1)^i (g_{i-1}- g_{i+1}),
\end{array}
\ee
\be
\begin{array}{lcl}
D_3g_i =g_i \gamma_i^+,&~~~~~~~~~~~~&
D_4g_i =(-1)^i g_i \gamma_i^+, \\
D_3\gamma_i^- =g_{i+1}- g_{i-1},&&
 D_4\gamma_i^- =(-1)^i ( g_{i+1}- g_{i-1}),\\
D_3\gamma_i^+ =\partial{\rm ln}g_i,&&
D_4\gamma_i^+ =-(-1)^i \partial{\rm ln}g_i.
\end{array}
\ee
Thus, equations \p{g-toda-1d} reproduce the 1D $N=4$ supersymmetric
Toda lattice equations.

In terms of the new variables (\ref{g-basis}) the automorphism $f$
(\ref{toda-auto}) becomes
\ba \label{g-toda-auto}
\gamma_j^-\ \stackrel{f}{\longmapsto}\ (-1)^j \gamma_j^+,\ \ \ \
\gamma_j^+\ \stackrel{f}{\longmapsto}\ -(-1)^j \gamma_j^-,
\ea
and it transforms the flows
(\ref{g-toda-1d-1order})--(\ref{g-toda-1d-1order-susy}) and Hamiltonians
(\ref{g-ham}) as follows:
\ba\label{g-toda-auto-ham}
&&\{\partial,D_1,D_2,D_3,D_4\}\ \stackrel{f}{\longmapsto}
\{\partial,D_4,D_3,-D_2,-D_1\},\nn\\
&&\{H_k^{N=4},S_{1,k}^{N=4},S_{2,k}^{N=4},S_{3,k}^{N=4},S_{4,k}^{N=4}\}\
 \stackrel{f}{\longmapsto}
\{H_k^{N=4},S_{4,k}^{N=4},S_{3,k}^{N=4},-S_{2,k}^{N=4},-S_{1,k}^{N=4}\}.
\ea

\subsection{Fermionic Hamiltonians}
\label{FerH}
The above-described 1D $N=4$ supersymmetric Toda lattice hierarchy is
a bi-Hamiltonian system,
and it includes both
bosonic and fermionic flows which are generated via bosonic
and fermionic Hamiltonians. The bosonic Hamiltonians are
produced by means of formula (\ref{str}), while the origin of the fermionic
Hamiltonians is rather mysterious so far. In this section, we
deduce general expressions generating fermionic Hamiltonians.

The $N=(2|2)$ Toda lattice equations (\ref{g-toda}) can be derived as a
subsystem of more general $N=(1|1)$ 2D supersymmetric Toda lattice (TL) hierarchy
defined via the following Lax pair representation \cite{KS2}:
\ba\label{SK-Lax-eq}
&&D^{\pm}_n \mathtt (L^{{\alpha}})^{m}_{*}= \mp {\alpha} (-1)^{nm}
\Bigr[{(((\mathtt L^{\pm})^{n}_{*})}_{-{\alpha} })^{*(m)},
\mathtt (L^{{\alpha}})^m_{*}\Bigl\}, \quad {\alpha} =+,-, \quad n,m \in {\mathbb N},
\nonumber\\
&&(\mathtt L^{{\alpha}})^{2m}_{*}\equiv \Bigl(~\frac{1}{2}~\Bigr[
(\mathtt L^{{\alpha}})^{*}, (\mathtt L^{{\alpha}})\Bigl\}~\Bigr)^m, \quad
(\mathtt L^{{\alpha}})^{2m+1}_{*}\equiv \mathtt L^{{\alpha}}
~(\mathtt L^{{\alpha}})^{2m}_{*}, \nonumber\\
&&\mathtt L^{+}=
\sum^{\infty}_{k=0} u_{k,j}e^{(1-k){\partial}}, \quad
u_{0,j}=1, \quad
\mathtt L^{-}=\sum^{\infty}_{k=0}
v_{k,j}e^{(k-1){\partial}}, \nonumber\\
&&(\mathtt L^{+})^*=
\sum^{\infty}_{k=0}(-1)^k u_{k,j}e^{(1-k){\partial}}, \quad
(\mathtt L^{-})^*=\sum^{\infty}_{k=0} (-1)^k
v_{k,j}e^{(k-1){\partial}}, \nonumber\\
&& d_{\mathtt (L^{\pm})^{2m+1}_{*}}=1,
\quad d_{\mathtt (L^{\pm})^{2m}_{*}}=0.
\ea
All details concerning the
$N=(1|1)$ 2DTL
hierarchy can be found in \cite{KS1,KS2}, here we only explain the notation.
The generalized graded bracket operation $[...,...\}$, entering into eqs.
\p{SK-Lax-eq},
on the space of operators ${\mathbb O}$
with the grading $d_{{\mathbb O}}$
and the involution $*$ is defined as
\begin{eqnarray}\label{SK-bracket}
\Bigl[ {\mathbb O}_1, {\mathbb O}_2 \Bigr\}:={\mathbb O}_1{\mathbb O}_2
 - (-1)^{d_{{\mathbb O}_1}d_{{\mathbb O}_2}}~{{\mathbb O}_2}^{*(d_{{\mathbb
O}_1})}~ {{\mathbb O}_1}^{*(d_{{\mathbb O}_2})}
\end{eqnarray}
where ${{\mathbb O}}^{*(m)}$  denotes the $m$-fold action of the
involution $*$ on the operator ${\mathbb O}$.
Equations (\ref{SK-Lax-eq}) are written for the
composite Lax operators
 \begin{eqnarray}
\label{SK-Lax-com}
 (\mathtt L^{+})^{m}_{*}=
\sum^{\infty}_{k=0} u^{(m)}_{k,j}e^{(m-k){\partial}}, \quad
u^{(m)}_{0,j}=1, \quad (\mathtt L^{-})^{m}_{*}=\sum^{\infty}_{k=0}
v^{(m)}_{k,j}e^{(k-m){\partial}}
\end{eqnarray}
where  $u^{(m)}_{k,j}$ and $v^{(m)}_{k,j}$ (with
$u^{(1)}_{k,j}\equiv u_{k,j},~v^{(1)}_{k,j}\equiv v_{k,j}$) are the functionals
of the original bosonic $u_{2k,j}$,$v_{2k,j}$ and fermionic
$u_{2k+1,j}$,$v_{2k+1,j}$ lattice fields which parameterize the Lax operators
$L^{\pm}$ \p{SK-Lax-eq}.
The operator
$e^{l{\partial}}$ ($l \in {\mathbb Z}$) acts on these fields as
the discrete lattice shift
\begin{eqnarray}
e^{l{\partial}} u^{(m)}_{k,j} \equiv u^{(m)}_{k,j+l}
e^{l{\partial}}, \quad e^{l{\partial}} v^{(m)}_{k,j} \equiv v^{(m)}_{k,j+l}
e^{l{\partial}},
\end{eqnarray}
and the subscript $+(-)$ in eqs. \p{SK-Lax-eq}
means the part of the corresponding operators which includes
the operators $e^{l\partial}$ at $l \ge 0 (l<0)$.
The explicit form for the functional
  $u^{(m)}_{k,j}$ and $v^{(m)}_{k,j}$ can be obtained
through the representation of the composite Lax operators
$(\mathtt L^{\pm})^{m}_{*}$ \p{SK-Lax-eq},
(\ref{SK-Lax-com}) in terms of the Lax operators $L^{\pm}$ \p{SK-Lax-eq}.
The fields $u_{k,j},\ v_{k,j}$ depend on the bosonic $t^\pm_{2n}$
and fermionic  $t^\pm_{2n+1}$ times, and $D_{2n}^\pm$ ($D_{2n+1}^\pm$)
in eq. (\ref{SK-Lax-eq})
means bosonic (fermionic) evolution derivatives
with the algebra
\begin{eqnarray}
[D^{+}_{n}~,~D^{-}_{l}\}=[D^{\pm}_{n}~,~D^{\pm}_{2l}]=0,\quad
\{D^{\pm}_{2n+1}~,~D^{\pm}_{2l+1}\}=2D^{\pm}_{2(n+l+1)}
\end{eqnarray}
which can be realized via
\begin{eqnarray}
D^{\pm}_{2n} ={\partial}^{\pm}_{2n}, \quad D^{\pm}_{2n+1}
={\partial}^{\pm}_{2n+1}+
\sum^{\infty}_{l=1}t^{\pm}_{2l-1}{\partial}^{\pm}_{2(k+l)},\quad
\partial^{\pm}_n={\frac{\partial}{\partial t^{\pm}_n}}.
\end{eqnarray}

Now using the above-described definitions one can derive flows
for the functionals $u_{k,j}^{(m)}$, $ v_{k,j}^{(m)}$ corresponding
to the Lax pair representation (\ref{SK-Lax-eq}).
Thus, we obtain \cite{KS1,KS2}
\begin{eqnarray}
\label{eq0}
D^{+}_n u^{(2m)}_{k,j}&=&\sum^{n}_{p=0}
(u^{(n)}_{p,j}u^{(2m)}_{k-p+n,j-p+n} \nonumber\\
&-&(-1)^{(p+n)(k-p+n)}u^{(n)}_{p,j-k+p-n+2m}u^{(2m)}_{k-p+n,j}),\\
\label{eq1}
D^{+}_{2n+1} u^{(2m+1)}_{k,j}&=&\sum^{k}_{p=1}
((-1)^{p+1}u^{(2n+1)}_{p+2n+1,j}u^{(2m+1)}_{k-p,j-p} \nonumber\\
&+&(-1)^{p(k-p)}u^{(2n+1)}_{p+2n+1,j-k+p+2m+1}u^{(2m+1)}_{k-p,j}),\\
\label{eq2}
D^{+}_{2n} u^{(2m+1)}_{k,j}&=&\sum^{2n}_{p=0}
((-1)^{p}u^{(2n)}_{p,j}u^{(2m+1)}_{k-p+2n,j-p+2n} \nonumber\\
&-&(-1)^{p(k-p)}u^{(2n)}_{p,j-k+p-2n+2m+1}u^{(2m+1)}_{k-p+2n,j}),\\
\label{eq3}
D^{-}_n u^{(m)}_{k,j}&=&\sum^{n-1}_{p=0}
((-1)^{(p+n)m}v^{(n)}_{p,j}u^{(m)}_{k+p-n,j+p-n} \nonumber\\
&-&(-1)^{(p+n)(k+p-n)}v^{(n)}_{p,j-k-p+n+m}u^{(m)}_{k+p-n,j}),\\
\label{eq4}
D^{+}_n v^{(m)}_{k,j}&=&\sum^{n}_{p=0}
((-1)^{(p+n)m}u^{(n)}_{p,j}v^{(m)}_{k+p-n,j-p+n} \nonumber\\
&-&(-1)^{(p+n)(k+p-n)}u^{(n)}_{p,j+k+p-n-m}v^{(m)}_{k+p-n,j}),\\
\label{eq5}
D^{-}_{2n} v^{(2m+1)}_{k,j}&=&\sum^{2n-1}_{p=0}
((-1)^{p}v^{(2n)}_{p,j}v^{(2m+1)}_{k-p+2n,j+p-2n} \nonumber\\
&-&(-1)^{p(k-p)}v^{(2n)}_{p,j+k-p+2n-2m-1}v^{(2m+1)}_{k-p+2n,j}),\\
\label{eq6}
D^{-}_{2n+1} v^{(2m+1)}_{k,j}&=&\sum^{k}_{p=0}
((-1)^{p+1}v^{(2n+1)}_{p+2n+1,j}v^{(2m+1)}_{k-p,j+p} \nonumber\\
&+&(-1)^{p(k-p)}v^{(2n+1)}_{p+2n+1,j+k-p-2m-1}v^{(2m+1)}_{k-p,j}),\\
\label{eq7}
D^{-}_n v^{(2m)}_{k,j}&=&\sum^{n-1}_{p=0}
(v^{(n)}_{p,j}v^{(2m)}_{k-p+n,j+p-n} \nonumber\\
&-&(-1)^{(p+n)(k-p+n)}v^{(n)}_{p,j+k-p+n-2m}v^{(2m)}_{k-p+n,j})
\end{eqnarray}
where in the right-hand side of these equations, all the fields
$\{u^{(m)}_{k,j},~v^{(m)}_{k,j}\}$ with $k < 0$  must be set equal to
zero.

The $N=(2|2)$ supersymmetric 2DTL equation belongs to the system of
equations (\ref{eq0})--(\ref{eq7}). In order to see that,
let us consider eqs. (\ref{eq3}) at $\{n=m=k=1\}$
\begin{eqnarray}\label{e1}
D^{-}_1 u_{1,j} = - v_{0,j} - v_{0,j+1}
\end{eqnarray}
and eqs. (\ref{eq4}) at $\{n=m=1, k=0\}$
\begin{eqnarray}\label{e2}
D^{+}_1 v_{0,j}= v_{0,j}(u_{1,j} - u_{1,j-1}).
\end{eqnarray}
Then, eliminating the field $u_{1,j}$ from eqs.
(\ref{e1})--(\ref{e2}) we obtain
\begin{eqnarray}\label{e3}
D^{+}_1D^{-}_1 \ln v_{0,j}= v_{0,j+1} - v_{0,j-1}.
\end{eqnarray}
Equation (\ref{e3}) reproduces the $N=(1|1)$ superfield form of the
$N=(2|2)$ superconformal 2DTL equation (\ref{g-toda})
(see, e.g., refs. \cite{EH,dls} and
references therein). Indeed, in the terms of the
superfield components
\begin{eqnarray}\label{comp-sub}
g_{j} \equiv v_{0,j}\Bigr|,
\quad {\gamma}^{\pm}_{j} \equiv ({ D}^{\pm}_1\ln v_{0,j})\Bigr|
\end{eqnarray}
where $g_j$ ($\gamma_j^\pm$) are the bosonic (fermionic)
fields and $|$ means the $t^{\pm}_1\rightarrow 0$ limit,
eq. (\ref{e3})
 coincides with (\ref{g-toda})
at $D_2^- \to -\partial_1$,
 $D_2^+ \to \partial_2$.

Now we define the supertrace of the operators $\mathbb O_m$
\begin{eqnarray}\label{O}
{\mathbb O}_m=\sum^{\infty}_{k=-\infty}
f^{(m)}_{k,j}e^{(k-m){\partial}}, \quad m\in {\mathbb Z},
\end{eqnarray}
parameterized by the bosonic (fermionic) lattice functions
$f^{(m)}_{2k,j}$ ($f^{(m)}_{2k+1,j}$)
as a sum of all their diagonal elements of the trivial shift
operator with $l=0$ $( e^{0\partial}=1)$ multiplied by
the factor $(-1)^j$
\ba
str \mathbb O=\sum\limits_{j=-\infty}^{\infty}(-1)^j f^{(m)}_{m,j}.
\ea
One can easily verify that the main property of supertraces
\begin{eqnarray}\label{suptr0}
str \Bigl[ {\mathbb O}_1, {\mathbb O}_2 \Bigr\}=0
\end{eqnarray}
is indeed satisfied for the case of the generalized graded bracket operation
\p{SK-bracket}. Using this definition of the supertrace and Lax pair
representation \p{SK-Lax-eq} one can easily obtain conserved Hamiltonians
of the $N=(1|1)$ 2DTL hierarchy
\ba\label{ham}
H^{\alpha}_m = str \mathtt (L^{{\alpha}})^{m}_{*}, \quad
D^{\pm}_n H^{\alpha}_m = 0,
\quad {\alpha} =+,-, \quad m \in {\mathbb N}.
\ea
From this formula it is obvious that all bosonic Hamiltonians corresponding to
even values of $m$ are trivial $H^{\alpha}_{2n}=0$ like a supertrace of the
generalized graded bracket operation, while fermionic Hamiltonians
at odd values of $m$ are not equal to zero $H^{\alpha}_{2n-1}\neq 0$ in general.
Using eqs. \p{SK-Lax-com} we obtain more explicit superfield formulae for the latter
 \begin{eqnarray}
\label{SK-Lax-exp}
H^{+}_{2n-1}=
\sum^{\infty}_{j=-\infty}(-1)^j u^{(2n-1)}_{2n-1,j}, \quad
H^{-}_{2n-1}=
\sum^{\infty}_{j=-\infty}(-1)^j v^{(2n-1)}_{2n-1,j}
\end{eqnarray}
which in terms of superfield components look like
\ba\label{s-int}
{s}^+_{n}(u)\equiv H^{+}_{2n-1}\Bigr|
= \sum\limits_{j=-\infty }^{\infty}(-1)^j u_{2n-1,j}^{(2n-1)}\Bigr|,
\ \ \ \ \ \
{ s}^-_{n}(v)\equiv H^{-}_{2n-1}\Bigr|
= \sum\limits_{j=-\infty }^{\infty}(-1)^j v_{2n-1,j}^{(2n-1)}\Bigr|.
\ea
The functionals $u^{(m)}_{m,j}$
and $v^{(m)}_{m,j}$ can be expressed in terms of the fields $v_{0,j}$ only
(for details see \cite{KS1}) and then, using eqs. (\ref{eq1}),
(\ref{eq3}), (\ref{eq6})
and eqs. (\ref{g-toda-1d-1order}),
in terms of the fields $(g_j,c_j,\gamma^\pm_j)$
in such a way that
$s^-_{m}(v)$ and $ s^+_{m}(u)$ become fermionic
integrals of motion for the
N=4 Toda lattice equations (\ref{g-toda-1d-1order}).

To understand better how formulae (\ref{s-int}) work,
we finish this section with the examples and reproduce all fermionic
Hamiltonians $S^{N=4}_{s,k}$ given by (\ref{g-ham}).

From (\ref{e2}) and  the component correspondence  (\ref{comp-sub})
it directly follows that
\ba
s^+_{1}(u)=\sum\limits_{j=-\infty}^{\infty}(-1)^j u_{1,j}\Bigr|=
1/2 \sum\limits_{j=-\infty}^{\infty} (-1)^j (D_1 {\rm ln} v_{0,j})\Bigr |=
1/2\sum\limits_{j=-\infty}^{\infty}(-1)^j\gamma^+_j=1/2 S_{3,1}^{N=4}
\ea
where $S_{3,1}^{N=4}$ is the fermionic integral in eqs. (\ref{g-ham}).

A more complicated problem is to obtain the next fermionic integral
$s^+_{2}(u)$. First, using eq. \p{SK-Lax-eq} one can find the explicit form
of the functional $u_3^{(3)}$
\ba\label{u33}
s^+_{2}(u)=\ \sum\limits_{j=-\infty}^{\infty}(-1)^j u^{(3)}_{3,j}\Bigr |=
3\sum\limits_{j=-\infty}^{\infty}(-1)^j(u_{3,j}+u_{2,j}(u_{1,j}-u_{1,j-1}))
\Bigr|.
\ea
Then, let us consider eq. ({\ref{eq1}) at $\{n=m=0,k=2\}$
and $\{n=m=0,k=1\}$,
\ba
\label{eq1-2}
D_1^+u_{2,j}&=&u_{2,j}(u_{1,j-1}-u_{1,j})-u_{3,j}+u_{3,j+1},\\
\label{eq1-1}
D_1^+u_{1,j}&=&u_{2,j}+u_{2,j+1},
\ea
respectively.
From eq. (\ref{eq1-2}) it follows that
\ba\label{sum-u3j}
\sum\limits_{j=-\infty}^{\infty}(-1)^ju_{3,j}=
-\frac12
\sum\limits_{j=-\infty}^{\infty}(-1)^j(D_1^+u_{2,j}+
u_{2,j}(u_{1,j}-u_{1,j-1}));
\ea
the consequence of eqs. (\ref{eq1-1}) and (\ref{e2}) is
\ba\label{u2j}
D_1^+(u_{1,j}-u_{1,j-1})=u_{2,j+1}-u_{2,j-1}=\partial^+_2{\rm ln}v_{0,j}
\ \ \to \ \ u_{2,j}=
\sum\limits_{k=0}^{\infty}\partial{\rm ln}v_{0,j-2k-1}
\ea
and, at last, from eq. ({\ref{e2}) one can find that
\ba\label{u1j}
u_{1,j}=
\sum\limits_{k=0}^{\infty}D^+_1{\rm ln}v_{0,j-k}.
\ea
Now it remains to substitute (\ref{sum-u3j})--(\ref{u1j}) into
(\ref{u33}) and
to reproduce
fermionic Hamiltonian $S_{3,2}^{N=4}=2/3 s^+_2(u)$ (\ref{g-ham})
using (\ref{g-toda-1d-1order}) and (\ref{comp-sub}).

Analogously, one can find that
\ba
S_{1,1}^{N=4}=-2 s^-_1(v), \ \ \ \ \ \
S_{1,2}^{N=4}=-2/3 s^-_2(v). \ \ \ \ \ \
\ea

The two remaining series of fermionic Hamiltonians in eqs. (\ref{g-ham})
can easily be derived from the obtained ones
if one applies the automorphism transformations (\ref{g-toda-auto})
\ba
S_{2,m}&=&S_{3,m}(\gamma^+_j\to -(-1)^j\gamma_j^-,
\gamma^-_j\to (-1)^j\gamma_j^+),\nn\\
S_{4,m}&=&-S_{1,m}(\gamma^+_j\to -(-1)^j\gamma_j^-,
\gamma^-_j\to (-1)^j\gamma_j^+).
\ea

\subsection{Transition to the canonical basis for the N=4 Toda lattice
equations}
\label{N4-TL-can}
Our next task is to rewrite the N=4 Toda lattice
equations (\ref{g-toda-1d-1order})
in a canonical basis where
these equations admit a Lagrangian formulation
that is important in connection with the quantization problem.

Let us introduce the new basis $\{x_j,p_j,\chi_j^+,\chi_j^-\}$ in the
phase space $\{g_j,c_j,\gamma_j^+,\gamma_j^-\}$
\ba\label{g-basis-canon}
&&g_j=ie^{x_j-x_{j-1}}, \ \ \ \ \ \
\gamma^+_j=\chi^-_j+(-1)^j\chi^+_{j-1},\nn\\
&&c_j=-(-1)^j p_j, \ \ \ \ \ \
\gamma^-_j=i(\chi^-_{j-1}-(-1)^j\chi^+_j)
 \ea
where $i$ is the imaginary unity and we suppose that the new fields
 go to zero at infinity
\ba\label{g-canon-boun}
\lim_{j\to\pm\infty} \{x_j,p_j,\chi_j^+,\chi_j^-\}=0.
\ea

In terms of the new coordinates the first Hamiltonian structure
(\ref{g-1PB}) becomes canonical
 \ba\label{22-1PB-canon}
\{x_i, p_j\}_1 =\delta_{i, j},
 \ \ \ \ \ \ \ \{\chi^-_i, \chi^+_j\}_1 = \delta_{i, j}
\ea
and the  Hamiltonians (\ref{g-ham}) take the following form:
\ba\label{H-can}
H_1&=&-\sum_{j=-\infty}^\infty p_j, \nn\\
H_2&=&\sum_{j=-\infty}^\infty (-1)^j (\frac12 p_j^2-e^{x_j-x_{j-2}}-
 e^{x_j-x_{j-1}} (\chi^-_{j-1}-(-1)^j
\chi^+_j)(\chi^-_j+(-1)^j \chi^+_{j-1})),\nn\\
S_{1,1}&=&i\sum_{j=-\infty}^\infty (-1)^j (\chi^-_{j-1}-(-1)^j\chi^+_j),
\ \ \ \
S_{2,1}\ =\ i\sum_{j=-\infty}^\infty (\chi^-_{j-1}-(-1)^j\chi^+_j),\nn\\
\ \ \ \
S_{3,1}&=&\sum_{j=-\infty}^\infty (-1)^j(\chi^-_j+(-1)^j\chi^+_{j-1}),
\ \ \ \
S_{4,1}\ =\ \sum_{j=-\infty}^\infty  (\chi^-_j+(-1)^j\chi^+_{j-1}).
\ea

The Hamiltonian $H_2$ generates the following equations via the first
Hamiltonian structure (\ref{22-1PB-canon})
\ba\label{22-canon}
  \partial x_j&=& -(-1)^j\ p_j,\nn\\
  \partial p_j& =&- (-1)^j \left(e^{x_j - x_{j - 2}} -
          e^{x_{j + 2} -x_j} +  e^{x_j - x_{j - 1}}
(\chi^-_{j-1}-(-1)^j \chi^+_j)(\chi^-_j+(-1)^j \chi^+_{j-1})\right.\nn\\
&&\left.
+  e^{x_{j + 1} - x_j}
(\chi^-_j+(-1)^j \chi^+_{j+1})(\chi^-_{j+1}-(-1)^j \chi^+_j)
\right),\nn\\
  \partial \chi_j^-&=&-\left (  e^{x_{j + 1} - x_j}(\chi^-_j+(-1)^j
\chi^+_{j+1}) +  e^{x_j - x_{j - 1}}(\chi^-_j+(-1)^j
  \chi^+_{j-1})\right),\nn\\
  \partial \chi_j^+&=&- (-1)^ j  \left( e^{x_j
- x_{j - 1}}(\chi^-_{j-1}-(-1)^j \chi^+_j) + e^{x_{j + 1} -
          x_j}(\chi^-_{j+1}-(-1)^j \chi^+_j)\right).
 \ea

Following the standard procedure one can derive the Lagrangian ${\cal L}$
and the action ${\cal S}$
\ba
{\cal S}&=&\int dt {\cal L}=\int dt [
\sum_{j=-\infty}^\infty p_j \frac{\partial}{\partial t}x_j+
\chi^-_j\frac{\partial}{\partial t}\chi_j^+-H_2]\nn\\
&=&\int dt \sum_{j=-\infty}^\infty [\frac12 (\frac{\partial}{\partial t}x_j)^2+
\chi^-_j\frac{\partial}{\partial t}\chi_j^+\nn\\
&&+
(-1)^j( e^{x_j-x_{j-2}}+
e^{x_j-x_{j-1}} (\chi^-_{j-1}-
(-1)^j \chi^+_j)(\chi^-_j+(-1)^j \chi^+_{j-1}))].
\ea
The variation of the action ${\cal S}$ with respect to the fields
$\{ x_j,\chi^-_j,\chi^+_j\}$ produces the equations of motion
(\ref{22-canon}) for them with reversed sign of time
($\partial \to -\frac{\partial}{\partial t}$)
where the momenta $p_j$
are replaced by $ (-1)^j \frac{\partial}{\partial t}x_j$.

One important remark is in order. There is no one-to-one correspondence
between the phase space bases $\{g_j,c_j,\gamma_j^+,\gamma_j^-\}$ and
$\{x_j,p_j,\chi_j^+,\chi_j^-\}$ (\ref{g-basis-canon}).
Transformation (\ref{g-basis-canon}) can rather be treated as a
reduction
of the primary phase space $\{g_j,c_j,\gamma_j^+,\gamma_j^-\}$
to the subspace $\{x_j,p_j,\chi_j^+,\chi_j^-\}$ with a smaller symmetry.
Indeed, the direct consequence of eqs.
(\ref{g-basis-canon})--(\ref{g-canon-boun}) is the following constraints
on the original fields
\ba\label{reduc-cond}
\prod_{k=-\infty}^{\infty}(-i g_k)=1,\ \ \ \
\sum_{k=-\infty}^{\infty}(\gamma_k^-
-i \gamma_k^+)=0, \ \ \ \
\sum_{k=-\infty}^{\infty}(-1)^k(\gamma_k^-
+i \gamma_k^+)=0.
\ea
Equations (\ref{reduc-cond}) restrict the phase space of
(\ref{g-toda-1d-1order}) and change the symmetry properties
of the latter.
The first manifestation of such a restriction is the fact that Hamiltonians
$H_1$ and $S_{n,1}$  (\ref{H-can})
no longer belong to the center of the first Hamiltonian
structure and generate the following nontrivial flows
via the algebra (\ref{22-1PB-canon}):
\ba
&& \partial_{H_1} x_j=1,\ \ \
D_{S_{1,1}}\chi_j^-=-i, \ \ \
D_{S_{1,1}}\chi_j^+=-(-1)^j i, \ \ \
D_{S_{2,1}}\chi_j^-=-(-1)^j i, \ \ \
D_{S_{2,1}}\chi_j^+=i, \nn\\
&&D_{S_{3,1}}\chi_j^-=1, \ \ \
D_{S_{3,1}}\chi_j^+=(-1)^j , \ \ \
D_{S_{4,1}}\chi_j^-=-(-1)^j , \ \ \
D_{S_{4,1}}\chi_j^+=1.
\ea
Furthermore, conditions (\ref{reduc-cond}) break the $N=4$ supersymmetry
transformations (\ref{g-toda-1d-1order-susy}) and in order to restore
the $N=4$ supersymmetry,
one needs to impose the following additional
constraints on the fields:
\ba
\sum_{k=-\infty}^{\infty}(-1)^k\gamma_k^\pm=0,\ \ \ \ \
\sum_{k=-\infty}^{\infty}\gamma_k^\pm=0,\ \ \ \ \
\sum_{k=-\infty}^{\infty}(-1)^k c_{k}=0.
\ea
However, the following $N=2$ supersymmery transformations in terms
of fermionic flows (\ref{g-toda-1d-1order-susy}):
\ba
\tilde D_1=i D_1+ D_3,\ \ \ \  \tilde D_2=i D_2- D_4,\ \ \ \ \ \ \
\tilde D_1^2=2\partial,\ \ \ \
\tilde D_2^2=-2\partial
\ea
are consistent with the constraints (\ref{reduc-cond}) and provide
the $N=2$ sypersymmetry for the infinite Toda lattice in the canonical basis
(\ref{22-canon})
\ba\label{22-canon-susy}
\tilde D_1 x_j&=&
\chi_j^-+(-1)^j \chi^+_j,\nn\\
\tilde D_1p_j&=&
e^{x_{j+1}-x_j}(\chi_{j+1}^+-\chi_j^++(-1)^j(\chi_{j}^-+\chi^-_{j+1}))\nn\\
&&+
e^{x_{j}-x_{j-1}}(\chi_{j-1}^+-\chi_{j}^++(-1)^j(\chi^-_{j}+\chi_{j-1}^-))\
,\nn\\
 \tilde D_1\chi^-_j&=&e^{x_{j+1}-x_j} -e^{x_{j}-x_{j-1}}-(-1)^j p_j\ ,\nn\\
 \tilde D_1\chi^+_j&=&(-1)^j (e^{x_{j}-x_{j-1}} -e^{x_{j+1}-x_{j}})-p_j,
  \nn\\
&&~\\
\tilde D_2 x_j&=&
\chi_j^+-(-1)^j \chi^-_j \ ,\nn\\
\tilde D_2p_j&=&
e^{x_{j+1}-x_j}(\chi_{j+1}^--\chi_j^--(-1)^j(\chi_{j}^++\chi^+_{j+1}))\nn\\
&&+
e^{x_{j}-x_{j-1}}(\chi_{j-1}^--\chi_{j}^--(-1)^j(\chi^+_{j}+\chi_{j-1}^+)),
\nn \\
 \tilde D_2\chi^+_j&=&e^{x_{j+1}-x_j} -e^{x_{j}-x_{j-1}}+(-1)^j p_j\ , \nn \\
 \tilde D_2\chi^-_j&=&(-1)^j (e^{x_{j+1}-x_{j}} -e^{x_{j}-x_{j-1}})-p_j.
  \nn
\ea

Equation (\ref{22-canon}) can be represented in the superfield form
\ba\label{22-canon-sf}
{\cal D}_+ {\cal D}_-\Phi_j= 2 (-1)^j (e^{\Phi_{j+1}-\Phi_{j}}-
e^{\Phi_{j}-\Phi_{j-1}})
\ea
where $\Phi_j$ is the bosonic $N=2$ superfield with the components
\ba
x_j=\Phi_j \Bigr |,\ \ \ \
 \chi_j^-+(-1)^j \chi^+_j={\cal D}_+ \Phi_j \Bigr |, \ \ \ \
\chi_j^+-(-1)^j \chi^-_j={\cal D}_- \Phi_j \Bigr|.
 \ea
Here $|$ means the
$\theta^\pm\to 0$ limit and ${\cal D}_\pm$ are the fermionic covariant
derivatives
\ba
{\cal D}_\pm= \frac{\partial}{\partial_{\theta^\pm}}\pm
2\theta^\pm\partial,\ \ \ \ \
       {\cal D}_\pm^2=\pm 2 \partial, \ \ \ \ \
\{{\cal D}_+,{\cal D}_-\}=0.
\ea

In order to rewrite the second Hamiltonian structure (\ref{g-2PB})
in terms of the new fields $\{x_j,p_j,\chi_j^+,\chi_j^-\}$, we invert
(\ref{g-basis-canon})
\ba \label{inv-can-22}
x_j&=& c \sum_{k=-\infty}^j ({\rm ln} g_k-i\pi/2) +(c-1)
\sum_{k=j+1}^\infty({\rm ln} g_k -i\pi/2 ),
\ \ \ \ \
p_j=-(-1)^j c_j,
\nn\\
\chi^+_j&=&
 (-1)^j\sum_{k=-\infty}^{-1} (c_+ \gamma^+_{j+2k+1}+ic_-\gamma^-_{j+2k+2})
+
 (-1)^j\sum_0^{\infty} ((c_+-1) \gamma^+_{j+2k+1}+i
(c_--1)\gamma^-_{j+2k+2}),\nn\\
\chi^-_j&=&
 \sum_{k=-\infty}^{-1} (c_+ \gamma^+_{j+2k+2}+ic_-\gamma^-_{j+2k+1})
-
 \sum_0^{\infty} ((c_+-1)\gamma^+_{j+2k+1}+i(c_--1)\gamma^-_{j+2k+1})
\ea
and find the Poisson brackets between the fields $\{x_j,p_j,\chi_j^+,
\chi_j^-\}$ using relations (\ref{g-2PB}). Equations (\ref{inv-can-22})
contain three arbitrary parameters $c$ and $c_\pm$. However, the second
Hamiltonian structure (\ref{g-2PB})
is not consistent with constraints
(\ref{reduc-cond}) in general,
and it is not guaranteed {\it a priori } that the Poisson
brackets obtained in such a way obey the Jacobi identities.
The test of the Jacobi identities shows that the Poisson brackets obtained
form a closed algebra only at $c=1$, $c_\pm=0$ and have the following
explicit form:
\ba\label{22-2PB-canon}
\{x_i,x_j\}_2&=&(-1)^j \delta^+_{i,j}-(-1)^i \delta^-_{i,j}\ ,\nn\\
\{x_i,p_j\}_2&=&- (-1)^j p_j \delta_{i,j}\ ,\nn\\
\{p_i,p_j\}_2&=&-(-1)^j\biggl(
 e^{x_i-x_j} (\chi_i^-+(-1)^i \chi^+_{i-1})(\chi_{i-1}^--(-1)^i
 \chi^+_i)\delta_{i,j+1}+e^{x_i-x_j}\delta_{i,j+2}\nn \\
&&+ e^{x_j-x_i} (\chi_j^-+(-1)^j \chi^+_{j-1})(\chi_{j-1}^--(-1)^j
 \chi^+_j)\delta_{i,j-1}
-e^{x_j-x_i}\delta_{i,j-2}\biggr)\ ,\nn\\
  \{p_i,\chi_j^\pm\}_2&=&
 \varrho_{i+j}^+ \left[e^{x_{i+1}-x_i}(\mp\chi^\pm_i+(-1)^i
 \chi^\mp_{i+1}) \delta_{i,j-1}^++e^{x_i-x_{i-1}}(\pm\chi^\pm_i-(-1)^i
 \chi^\mp_{i-1}) \delta_{i,j+1}^+\right]\nn\\
&& + \varrho_{i+j}^- \left[e^{x_i-x_{i-1}}(\pm\chi^\pm_{i-1}+(-1)^i
 \chi^\mp_i) \delta_{i,j+1}^++e^{x_{i+1}-x_i}(\mp\chi^\pm_{i+1}-(-1)^i
 \chi^\mp_i)\delta_{i,j-2}^+\right]\ ,\nn\\
\{x_i,\chi_j^\pm\}_2&=&
\varrho^+_{i+j}  \biggl[ (-1)^j(\chi_j^\pm-\chi_i^\pm)+
2\sum_{k=1}^{(i-j)/2}(\mp\chi^\pm_{i-2k+1}\pm
(-1)^j\chi^\pm_{j+2k})\biggr]\delta_{i,j+1}^+\nn\\
&& +\varrho^-_{i+j} \biggl[ (\mp\chi_i^\mp+(-1)^j
\chi^\pm_j)\delta_{i,j}^+
+ 2 \sum_{k=1}^{(i-j-1)/2}(\mp \chi^\mp_{i-2k}+
(-1)^j \chi^\pm_{j+2k})\delta_{i,j+1}^+\biggr]
\ ,\nn\\
\{\chi^-_i,\chi^+_j\}_2&=&\varrho^+_{i+j}\biggl[ 2(-1)^j \delta_{i,j}
 \sum_{k=i+1}^\infty p_k+
\biggl(  e^{x_i-x_{i-1}}+
e^{x_{i+1}-x_i}-(-1)^j(p_i+2  \sum_{k=i+1}^\infty p_k)\biggr)\delta^+_{i,j}
 \nn\\
&& +\biggl(  e^{x_j-x_{j-1}}+
e^{x_{j+1}-x_j}+(-1)^j(p_j+2  \sum_{k=j+1}^\infty p_k)\biggr)\delta^-_{i,j}
\biggr ]\ ,\nn\\
\{\chi^\pm_i,\chi^\pm_j\}_2&=&\varrho^-_{i+j}\biggl[
\biggl(- (-1)^i  e^{x_{i+1}-x_i}\mp p_i\mp 2\sum_{k=i+1}^\infty
p_k\biggr) \delta_{i,j}^+
- (-1)^i e^{x_{i}-x_{i-1}}\delta_{i,j+1}^+  \nn\\
&& +\biggl(- (-1)^j
e^{x_{j+1}-x_j}\mp p_j\mp 2\sum_{k=j+1}^\infty p_k\biggr) \delta_{i,j}^-
 - (-1)^j e^{x_{j}-x_{j-1}}\delta_{i,j-1}^-  \biggr]
 \ea
where we have introduced the notation
$$
\varrho^\pm_j =(1\pm(-1)^j)/2, \quad
\delta_{i,j}^+ =\left\{
   \begin{array}{rcl}
1,& \mbox{if} &i>j\\
0,& \mbox{if} &i\leq j \\
 \end{array}
\right.
\ ,
\hspace*{1cm}
\mbox{}
\delta_{i,j}^- =\left\{
   \begin{array}{rcl}
1,& \mbox{if} &i<j\\
0,& \mbox{if} &i\geq j \\
 \end{array}
\right.
$$
with the property
$$\delta_{i,j}^-+\delta_{i,j}^++\delta_{i,j}\equiv 1.$$
One can check that the Hamiltonian $H_1$ (\ref{H-can})
reproduces equations (\ref{22-canon})
via the second Hamiltonian structure (\ref{22-2PB-canon}).

\section{Reduction: 1D N=2 supersymmetric Toda lattice hierarchy}
\setcounter{equation}{0}
\label{N2-TL}
\subsection{Bi-Hamiltonian structure of the 1D N=2 Toda lattice hierarchy}
The 1D $N=2$ Toda lattice hierarchy was proposed and studied in detail in \cite{BS}.
In this section, we reproduce its bi-Hamiltonian description \cite{BS}
reducing the bi-Hamiltonian structure \p{1pb-TODA}--\p{2pb-TODA} of the 1D
generalized fermionic Toda lattice
hierarchy by reduction constraint \p{reds1}.

Our starting point is eqs. (\ref{ab-toda})
with boundary conditions
$I)$ and $III)$ (\ref{TODA-bound}) in one-dimensional space \p{red-one}.
Substituting reduction constraint \p{reds1}
into (\ref{ab-toda}), we obtain the following equation for the fields $\bar a_j$
\ba\label{a-bar-eq}
\partial \bar a_j=\beta_j\bar\alpha_j+\alpha_j\bar\beta_j
\ea
which can easily be solved. Here, we note that the system
(\ref{ab-toda})
is scale-invariant and length dimensions of the involved fields are:
$[b_j]=[\bar b_j]=-2$, $[a_j]=[\bar a_j]=-1$,
 $[\beta_j]=[\bar \beta_j]=-3/2$,
$[\alpha_j]=[\bar \alpha_j]=-1/2$.
Keeping this in mind we obtain the scale-invariant solution
to eq. (\ref{a-bar-eq})
\be\label{ab-sol}
\bar a_j=-\frac{\beta_j\bar \beta_j}{b_j}.
\ee
Substituting this solution into eqs. (\ref{ab-toda}) at
$\partial_2=\partial_1=\partial$
we arrive at the following equations:
\begin{eqnarray}  \label{ab-toda-red}
&&\partial b_j=b_j(a_j-a_{j-1}),\ \ \
\partial a_j=
b_{j+1}-b_j+\beta_j\bar\alpha_j+ \alpha_{j+1}\bar\beta_{j+1},\nn \\
&& \partial\beta_j=a_j\beta_j -b_j \alpha_j,\ \ \
\partial \bar\beta_j=-a_{j-1}\bar \beta_j-b_j\bar\alpha_j, \nn \\
&&\partial \alpha_j=\beta_j-\beta_{j-1},\ \ \
\partial \bar\alpha_j=\bar\beta_j-\bar\beta_{j+1}.
\end{eqnarray}
Fermionic symmetries (\ref{toda-ab-susy-12}) become inconsistent
after reduction $\bar b_j=0$ because in this case the fields $e_j$
(\ref{e-repr})
become singular.
As concerns fermionic symmetries (\ref{toda-ab-susy-34}), they are consistent
and take the following form:
\ba \label{ab-toda-red-susy}
\begin{array}{lcl}
D_1b_j=b_j(\alpha_j-\bar \alpha_j),&~~~~~~~~&
D_2b_j=b_j(\alpha_j+\bar \alpha_j),\\
D_1a_j=\bar\beta_{j+1}+\beta_j,&&
D_2a_j=-\bar\beta_{j+1}+\beta_j,\\
D_1\beta_j=-b_j+\beta_j\bar\alpha_j,&&
D_2\beta_j=b_j-\beta_j\bar\alpha_j,\\
D_1\bar\beta_j=b_j-\bar\beta_j\alpha_j,&&
D_2\bar\beta_j=b_j-\bar\beta_j\alpha_j,\\
D_1\alpha_j=-a_{j-1}-{\displaystyle
\frac{\beta_j\bar\beta_j}{b_j}},&&
D_2\alpha_j=a_{j-1}+{\displaystyle
\frac{\beta_j\bar\beta_j}{b_j}},\\
D_1\bar\alpha_j=-a_j-{\displaystyle
\frac{\beta_j\bar\beta_j}{b_j}},&&
D_2\bar\alpha_j=-a_j-{\displaystyle
\frac{\beta_j\bar\beta_j}{b_j}}.
\end{array}
\ea

The system (\ref{ab-toda-red}) is supplied
with the boundary conditions $I)$ and $III)$ (\ref{TODA-bound}).
Let us recall that for the boundary conditions $III)$ (\ref{TODA-bound})
the system (\ref{ab-toda}) does not possess any supersymmetry (see
the paragraph with eqs. \p{bc}).
In terms of fields (\ref{ab-basis}) the boundary conditions
$I)$ and $III)$ (\ref{TODA-bound}) are
\ba\label{ab-bc}
Ib).&&
\lim_{j\to\pm\infty}\{b_j,a_j,\alpha_j,\bar\alpha_j,
\beta_j,\bar\beta_j\}=0,\nn \\
IIb).
&&\lim_{j\to\pm\infty}b_j=1,\ \ \
\lim_{j\to\pm\infty}\{a_j,\alpha_j,\bar\alpha_j,
\beta_j,\bar\beta_j\}=0,
\ea
respectively. Therefore, we  conclude  that
system (\ref{ab-toda-red}) possesses $N=2$ supersymmetry only for the
boundary conditions $Ib)$ \p{ab-bc}, while
for the boundary conditions $IIb)$ \p{ab-bc}
it is not supersymmetric.

The first (\ref{1pb-TODA}) and second (\ref{2pb-TODA}) Hamiltonian
structures in the basis
$\{b_j,\bar b_j,a_j,\bar a_j,\alpha_j,\bar\alpha_j,\beta_j,\bar\beta_j\}$
(\ref{ab-basis}) look like
\ba\label{1pb-ab-basis}
\{b_i,a_j\}_1&=&b_i (\delta_{i,j}-\delta_{i,j+1}),\nn\\
\{\bar b_i,\bar a_j\}_1&=&\bar b_i (\delta_{i,j+1}-\delta_{i,j}),\nn\\
\{a_i,\beta_j\}_1&=&-\beta_j \delta_{i,j},\nn\\
\{\bar a_i,\bar\beta_j\}_1&=&\bar\beta_j \delta_{i,j},\nn\\
\{\bar a_i,\beta_j\}_1&=&-\beta_j \delta_{i,j},\nn\\
\{ a_i,\bar\beta_j\}_1&=&\bar\beta_j \delta_{i,j-1},\nn\\
\{ \beta_i,\bar\beta_j\}_1&=&b_j
                      \delta_{i,j}-\bar b_j\delta_{i,j-1},\nn\\
\{ \alpha_i,\bar\alpha_j\}_1&=&
                      \delta_{i,j}-\delta_{i,j+1}
\ea
and
\ba\label{2pb-ab-basis}
\begin{array}{lcl}
\{b_i,b_j\}_2=-b_i b_j (\delta_{i,j+1}-\delta_{i,j-1}),&~~~~~&
\{\bar b_i,\bar b_j\}_2=\bar b_i \bar b_j (\delta_{i,j+1}-\delta_{i,j-1}),\\
\{a_i,a_j\}_2=-(b_i \delta_{i,j+1}-b_j \delta_{i,j-1}),&&
\{\bar a_i,\bar a_j\}_2=\bar b_i \delta_{i,j+1}-\bar b_j \delta_{i,j-1},\\
\{a_i,\bar a_j\}_2=\bar \alpha_j \beta_i \delta_{i,j}-\alpha_j \bar\beta_j
\delta_{i,j-1},&&
\{b_i,a_j\}_2=-b_i a_j (\delta_{i,j+1}-\delta_{i,j}),\\
\{\bar b_i,\bar a_j\}_2=\bar b_i \bar a_j (\delta_{i,j+1}-\delta_{i,j}),&&
\{b_i,\alpha_j\}_2=-b_i \alpha_j \delta_{i,j},\\
\{\bar b_i,\bar \alpha_j\}_2=\bar b_i \bar \alpha_j \delta_{i,j+1},&&
\{b_i,\bar \alpha_j\}_2=b_i \bar \alpha_j \delta_{i,j},\\
\{\bar b_i,\alpha_j\}_2=-\bar b_i \alpha_j \delta_{i,j},&&
\{b_i,\beta_j\}_2=-b_i \beta_j\delta_{i,j+1},\\
\{\bar b_i,\bar \beta_j\}_2=\bar b_i \bar \beta_j\delta_{i,j+1},&&
\{\bar b_i, \beta_j\}_2=-\bar b_i  \beta_j\delta_{i,j},\\
\{ b_i,\bar \beta_j\}_2= b_i \bar \beta_j\delta_{i,j-1},&&
\{a_i,\beta_j\}_2=-a_i\beta_j\delta_{i,j}+b_j \alpha_j \delta_{i,j-1},\\
\{\bar a_i,\bar \beta_j\}_2=\bar a_i\bar\beta_j\delta_{i,j}+\bar b_j
\bar\alpha_i \delta_{i,j-1},&&
\{ a_i,\bar \beta_j\}_2= a_i\bar\beta_j\delta_{i,j-1}+ b_i
\bar\alpha_j \delta_{i,j},\\
\{\bar a_i, \beta_j\}_2=-\bar a_i\beta_j\delta_{i,j}+\bar b_i
\alpha_i \delta_{i,j+1},&&
 \{\bar\beta_i, \alpha_j\}_2=-(\bar\beta_i\alpha_j\delta_{i,j}
+\bar b_i\delta_{i,j+1}-  b_i \delta_{i,j-1}),\\
 \{\bar a_i,\bar \alpha_j\}_2=-\bar \beta_i\delta_{i,j+1},&&
\{\alpha_i,\bar\alpha_j\}_2=-a_j\delta_{i,j+1}+\bar a_j\delta_{i,j},\\
 \{\bar a_i, \alpha_j\}_2=- \beta_i\delta_{i,j-1},&&
 \{\beta_i,\bar \alpha_j\}_2=\beta_i\bar\alpha_j\delta_{i,j} -b_i\delta_{i,j+1}+
                \bar b_j \delta_{i,j-1},\\
 \{a_i,\alpha_j\}_2=-\beta_i\delta_{i,j},&&
\{\bar\beta_i,\beta_j\}_2=(\beta_i\bar\beta_j-\bar b_j a_i)\delta_{i,j-1}
             +(\beta_i \bar\beta_j+b_i\bar a_j)\delta_{i,j},\\
 \{ a_i,\bar \alpha_j\}_2=-\bar \beta_j\delta_{i,j-1},&&
\end{array}
\ea
respectively.
One can easily see that the algebras (\ref{1pb-ab-basis}) and
(\ref{2pb-ab-basis}) are consistent with
the reduction constraints
\p{reds1} and (\ref{ab-sol}), so the 1D $N=2$ supersymmetric Toda lattice
equations (\ref{ab-toda-red})
can be represented as a bi-Hamiltonian system with the first Hamiltonian
structure
\ba\label{1pb-ab-basis-red}
\{b_i,a_j\}_1&=&b_i (\delta_{i,j}-\delta_{i,j+1}),\nn\\
\{a_i,\beta_j\}_1&=&-\beta_j \delta_{i,j},\nn\\
\{ a_i,\bar\beta_j\}_1&=&\bar\beta_j \delta_{i,j-1},\nn\\
\{ \beta_i,\bar\beta_j\}_1&=&b_j\delta_{i,j},\nn\\
\{ \alpha_i,\bar\alpha_j\}_1&=& \delta_{i,j}-\delta_{i,j+1}
\ea
and the second Hamiltonian structure
\ba\label{2pb-ab-basis-red}
\{b_i,b_j\}_2&=&-b_i b_j (\delta_{i,j+1}-\delta_{i,j-1}),\nn\\
\{b_i,a_j\}_2&=&-b_i a_j (\delta_{i,j+1}-\delta_{i,j}),\nn\\
\{a_i,a_j\}_2&=&-b_i \delta_{i,j+1}+b_j \delta_{i,j-1},\nn\\
\{b_i,\alpha_j\}_2&=&-b_i \alpha_j \delta_{i,j},\nn\\
\{b_i,\bar \alpha_j\}_2&=&b_i \bar \alpha_j \delta_{i,j},\nn\\
\{b_i,\beta_j\}_2&=&-b_i \beta_j\delta_{i,j+1},\nn\\
\{ b_i,\bar \beta_j\}_2&=& b_i \bar \beta_j\delta_{i,j-1},\nn\\
\{a_i,\beta_j\}_2&=&-a_i\beta_j\delta_{i,j}+b_j \alpha_j \delta_{i,j-1},\nn\\
\{ a_i,\bar \beta_j\}_2&=& a_i\bar\beta_j\delta_{i,j-1}+ b_i
\bar\alpha_j \delta_{i,j},\nn\\
\{a_i,\alpha_j\}_2&=&-\beta_i\delta_{i,j},\nn\\
 \{ a_i,\bar \alpha_j\}_2&=&-\bar \beta_j\delta_{i,j-1},\nn\\
 \{\bar\beta_i, \alpha_j\}_2&=&-\bar\beta_i\alpha_j\delta_{i,j}
+  b_i \delta_{i,j-1},\nn\\
 \{\beta_i,\bar \alpha_j\}_2&=&\beta_i\bar\alpha_j\delta_{i,j}
-b_i\delta_{i,j+1},\nn\\
\{\beta_i,\bar\beta_j\}_2&=&\beta_i\bar\beta_j\delta_{i,j-1},\nn\\
\{\alpha_i,\bar\alpha_j\}_2&=&-a_j\delta_{i,j+1}-\frac{\beta_j\bar \beta_j}{b_j}
\delta_{i,j}
\ea
where when calculating we have substituted the reduction constraints
\p{reds1} and (\ref{ab-sol}) into the original algebras (\ref{1pb-ab-basis})
and (\ref{2pb-ab-basis}).
Let us also present a few first bosonic and fermionic Hamiltonians of the
1D $N=2$ supersymmetric Toda lattice hierarchy obtained from Hamiltonians
\p{HAM-bos} and \p{HAM-fer} using reduction constraints \p{reds1} and (\ref{ab-sol})
\ba\label{ab-ham}
&&
H_1^{N=2}=-\sum\limits_{j=-\infty}^{\infty}(a_j+\frac{\beta_j\bar\beta_j}{b_j}),
\ \ \ \ \
H_2^{N=2}=-\sum\limits_{j=-\infty}^{\infty}(\frac12
a_j^2+b_j+\alpha_j\bar\beta_j -\beta_j\bar\alpha_j), \nn \\
&&
S_{1,1}^{N=2}=\sum\limits_{j=-\infty}^{\infty}(\alpha_j+\bar\alpha_j),\ \ \ \ \
S_{2,1}^{N=2}=\sum\limits_{j=-\infty}^{\infty}(\alpha_j-\bar\alpha_j),\nn\\
&&
S_{1,2}^{N=2}=\sum\limits_{j=-\infty}^{\infty}\left(\beta_j-\bar\beta_j-
\bar\alpha_j\frac{\beta_j\bar\beta_j}{b_j}+
(\alpha_j-\bar\alpha_j)
\sum\limits_{k=-\infty}^{j-1}(a_k+\frac{\beta_k\bar\beta_k}{b_k})
\right),\nn\\
&&
S_{2,2}^{N=2}=\sum\limits_{j=-\infty}^{\infty}\left(\beta_j+\bar\beta_j+
\bar\alpha_j\frac{\beta_j\bar\beta_j}{b_j}+
(\alpha_j+\bar\alpha_j)
\sum\limits_{k=-\infty}^{j-1}(a_k+\frac{\beta_k\bar\beta_k}{b_k})
\right).
\ea

\subsection{Transition to the canonical basis for the 1D N=2 Toda
lattice equations}
\label{N2-TL-can}
The transition to the canonical basis for the 1D $N=2$ supersymmetric
Toda lattice equations
(\ref{ab-toda-red}) with non-periodic boundary conditions
is possible only for the boundary conditions $IIb)$
(\ref{ab-bc}). As we have already mentioned, in this case the system
(\ref{ab-toda-red}) is not supersymmetric, nevertheless it can serve
as a basement for the building of the periodic $N=2$ Toda lattice
equations in the canonical basis.
In this section, we briefly discuss the representation of the
system (\ref{ab-toda-red}) in the canonical basis.

Following paper \cite{BS}, let us introduce the new basis
$\{x_j,p_j,\xi_j,\bar\xi_j,\eta_j,\bar\eta_j\} $
in the phase space
 $\{a_j,b_j,\alpha_j,\bar\alpha_j,\beta_j,\bar\beta_j\} $
\ba\label{ab-can}
&&  a_i = p_i,\ \ \ \
  b_i= e^{x_i - x_{i - 1}},\nn \\
&&  \beta_i= e^{x_i}  \xi_i, \ \ \ \
  \bar\beta_i= e^{-x_{i - 1}}\bar\xi_i,\nn \\
&&  \bar\alpha_i=-\bar\eta_i,\ \ \ \
  \alpha_i = \eta_{i - 1} - \eta_i
\ea
with the zero boundary conditions at infinity
\ba\label{ab-canon-boun}
\lim_{j\to\pm\infty} \{x_j,p_j,\xi_j,\bar\xi_j,\eta_j,\bar\eta_j\}=0.
\ea

In this basis the first Hamiltonian structure becomes canonical
\ba\label{ab-1pb-can}
  \{x_i, p_j\}_1 = \delta_{i, j}, \ \ \ \
  \{\xi_i, \bar\xi_j\}_1 = \delta_{i, j},\ \ \ \
  \{\eta_i,\bar\eta_j\}_1 =\delta_{i, j},
\ea
while the second Hamiltonian structure takes a more complicated form
\begin{eqnarray}\label{ab-2pb-can}
\{x_i, x_j\}_2& = & \delta_{i, j}^- -  \delta_{i, j}^+\ , \nn \\
  \{x_i, p_j\}_2& =&p_j \delta_{i, j}\ ,  \nn \\
  \{p_i, p_j\}_2& = &
      e^{x_j - x_i}\delta_{i, j - 1}  -
        e^{x_i - x_j}\delta_{i,j + 1}\ , \nn \\
  \{p_i,\xi_j \}_2& =& e^{-x_i} (\eta_i - \eta_j)\delta_{i,j - 1}\ , \nn\\
\{p_i, \eta_j\}_2& =& e^{x_i} \xi_i (
                 \tilde c -\delta_{i, j}^+)\ , \nn\\
  \{x_i, \xi_j\}_2& =& \xi_j(1-c-\delta_{i, j}^- )\ ,\nn \\
  \{x_i, \eta_j\}_2& =& \eta_i (1-\tilde c-\delta_{i, j}^-)
       + \eta_j( \delta_{i, j}^- -c)\ , \nn \\
  \{\xi_i,\eta_j\}_2 &=& \xi_i \eta_i(\tilde c-\delta_{i, j}^+ ) + \xi_i
          \eta_j(c - 1+\delta_{i,j}^+) \ , \nn \\
  \{p_i, \bar\eta_j\}_2& =& e^ {-x_i} \bar\xi_j \delta_{i, j - 1}\ , \nn \\
  \{p_i, \bar\xi_j\}_2& =&- e^{x_i} \bar\eta_j \delta_{i, j}\ , \nn \\
\{x_i, \bar\eta_j\}_2& =& \bar\eta_j (c-\delta_{i, j}^- )\  , \nn\\
\{x_i, \bar\xi_j\}_2& =& \bar\xi_j ( c-1+\delta_{i,j}^-)\ , \nn\\
 \{\bar\xi_i, \bar\eta_j\}_2& =& \bar\xi_i
    \bar\eta_j(c-1 + \delta_{i, j}^+ )\ ,\nn\\
 \{\xi_i,\bar\eta_j\}_2& =& \xi_i \bar\eta_j(1-c- \delta_{i, j}^+)
 + e^{-x_j}\delta_{i, j + 1}\ ,\nn\\
\{\eta_i, \bar\eta_j\}_2& = &
  \xi_j \bar\xi_j ( \delta_{i, j}^--\tilde c)
+ p_j (1-\tilde c-\delta_{i,j}^+) \, \nn\\
\{\bar\xi_i, \eta_j\}_2& =& (\bar\xi_i \eta_i+e^ {x_i})
(1-\tilde c-\delta_{i, j}^- )+ \bar\xi_i
\eta_j ( \delta_{i, j}^--c), \,
\end{eqnarray}
where $c$ is an arbitrary parameter and
$\tilde c=1$ or $0$. One can trace the origin of these
parameters if one
writes down the most general form of the inverse
transformations (\ref{ab-can})
\ba
&&x_i=c\sum_{k=-\infty}^i {\rm ln}b_k+
(c-1)\sum_{k=i+1}^{\infty} {\rm ln}b_k,\ \ \ \
\eta_i=-\tilde
c\sum_{k=-\infty}^i \alpha_k+ (1-\tilde c)\sum_{k=i+1}^{\infty} \alpha_k,\nn\\
&&p_i=a_i,\ \ \ \xi_i=e^{-x_i}\beta_i,\ \ \
\bar \xi_i=e^{x_{i-1}}\bar \beta_i, \ \ \ \bar\eta_i=\bar\alpha_i.
\ea
From the Jacobi identities one can fix
the parameter $\tilde c$ to be 1 or 0,
while the second parameter $c$ is left arbitrary.

In the canonical basis (\ref{ab-can}) the bosonic
Hamiltonians (\ref{ab-ham}) become
\begin{eqnarray}
H_1= - \sum_{i = 1}^n(p_i+\xi_i\bar\xi_i), \ \ \ \ \
 H_2=- \sum_{i = 1}^n(\frac12 p_i^2  + e^{x_i - x_{i - 1}} +
          e^{-x_i}\bar\xi_{i + 1}(\eta_{i + 1} -
                \eta_i ) + e^{x_i}\xi_i \bar\eta_i ),
\end{eqnarray}
and they generate, via the first (\ref{ab-1pb-can}) and second
(\ref{ab-2pb-can}) Hamiltonian structures, the following
equations \cite{BS}:
\begin{eqnarray}\label{ab-toda-can}
&&  \partial x_i= p_i, \ \ \
\partial\bar\xi_i= e^{x_i}\bar\eta_i, \ \ \
  \partial\xi_i=
    e^{-x_{i - 1}}(\eta_i -\eta_{i-1}),\nn \\
&&  \partial p_i= e^{x_{i + 1} - x_i} - e^{x_i - x_{i - 1}} -
      e^{x_i}\xi_i \bar\eta_i-
      e^{-x_i}
        \bar\xi_{i + 1} (\eta_i - \eta_{i + 1}),\nn \\
&&\partial \eta_i=
   - e^{x_i} \xi_i, \ \ \ \
\partial\bar \eta_i= e^{-x_i}\bar\xi_{i + 1} -
      e^{-x_{i - 1}} \bar\xi_i.
\end{eqnarray}
The parameters $c,\tilde c$ do not affect
equations (\ref{ab-toda-can}) via the second Hamiltonian structure
(\ref{ab-2pb-can}).

Let us also present the Lagrangian ${\cal L}$
and the action ${\cal S}$
\ba
{\cal S}&=&\int dt {\cal L}=\int dt [
\sum_{j=-\infty}^\infty p_j \frac{\partial}{\partial t}x_j+
\xi_j\frac{\partial}{\partial t}\bar\xi_j+
\eta_j\frac{\partial}{\partial t}\bar\eta_j
-H_2]\nn\\
&=&\int dt \sum_{j=-\infty}^\infty
[-\frac12 (\frac{\partial}{\partial t}x_j)^2+
\xi_j\frac{\partial}{\partial t}\bar\xi_j+
\eta_j\frac{\partial}{\partial t}\bar\eta_j\nn\\
&&
+ e^{x_i - x_{i - 1}} +
          e^{-x_i}\bar\xi_{i + 1}(\eta_{i + 1} -
                \eta_i ) + e^{x_i}\xi_i \bar\eta_i ].
\ea
One can easily verify that
the variation of the action ${\cal S}$ with respect to the fields
$\{ x_j,\xi_j,\bar\xi_j,\eta_j,\bar\eta_j\}$ produces equations of
motion (\ref{ab-toda-can}) for them with reversed sign of time
($\partial \to -\frac{\partial}{\partial t}$)
where the momenta $p_j$ are replaced by
$ -\frac{\partial}{\partial t}x_j$.

\section{Periodic Toda lattice hierarchies}
\setcounter{equation}{0}
\subsection{Periodic 2D generalized fermionic Toda lattice equations}
\label{perTL}
The $n$-periodic 2D generalized fermionic
 Toda lattice equations (\ref{TODA}) are characterized by the
 boundary conditions $IV)$ (\ref{TODA-bound}).
This system has completely  different symmetry
properties for odd and even values of the period $n$.
From now on we concentrate on the case with even value $n=2m$ of the period.

The $2m$-periodic 2D generalized fermionic
Toda lattice equations (\ref{TODA}) admit the zero-curvature representation
\be \label{Ln-eqn}
[\partial_1+L_{2m}^-,\partial_2 -L_{2m}^+]=0
\ee
with the $2m\times 2m$ matrices $L^{\pm}_{2m}$
\begin{eqnarray}\label{Ln-pair}
(L_{2m}^-)_{i,j}&=&\rho_i \delta_{i,j+1} +d_i \delta_{i,j+2}+ w^{-1}(\rho_1
\delta_{i,1}\delta_{j,2m} +d_1 \delta_{i,1}\delta_{j,2m-1} +d_{2}
\delta_{i,2}\delta_{j,2m}), \nn \\
(L_{2m}^+)_{i,j}&=&\delta_{i,j-2}+\gamma_i
\delta_{i,j-1}+c_i \delta_{i,j}+ w (\delta_{i,2m-1}\delta_{j,1}+
\delta_{i,2m}\delta_{j,2}+
\gamma_{2m} \delta_{i,2m}\delta_{j,1}),
\end{eqnarray}
\ba
 L_{2m}^-&=&\left(
\begin{array}{ccccccccc}
0&0&0&\dots&0&0&0&d_1/w&\rho_1/w\\
\rho_2&0&0&\dots&0&0&0&0&d_2/w\\
d_{3}&\rho_{3}&0&\dots&0&0&0&0&0\\
0&d_{4}&\rho_{4}&\dots&0&0&0&0&0\\
&&&\dots&\dots&&&&\\
&&&\dots&\dots&&&&\\
0&0&0&\dots&d_{2m-2}&\rho_{2m-2}&0&0&0\\
0&0&0&\dots&0&d_{2m-1}&\rho_{2m-1}&0&0\\
0&0&0&\dots&0&0&d_{2m}&\rho_{2m}&0\\
\end{array}
\right),\nn\\
&&
~\nn \\
 L^+_{2m}&=&\left(
\begin{array}{ccccccccc}
c_1&\gamma_1&1&0&\dots&0&0&0&0\\
0&c_2&\gamma_2&1&\dots&0&0&0&0\\
0&0&c_3&\gamma_3&\dots&0&0&0&0\\
0&0&0&c_4&\dots&0&0&0&0\\
&&&&\dots&\dots&&&\\
&&&&\dots&\dots&&&\\
0&0&0&0&\dots&0&c_{2m-2}&\gamma_{2m-2}&1\\
w&0&0&0&\dots&0&0&c_{2m-1}&\gamma_{2m-1}\\
w \gamma_{2m}&w&0&0&\dots&0&0&0&c_{2m}\\
\end{array}
\right) \nn
\ea
where $w$ is the spectral parameter of length dimension $[w]=-m$.

Now, following paper \cite{KulSk} let us give some definitions concerning
supermatrices.
For any $n\times n$ supermatrix  $F$ one can define the Grassmann parity
of rows and columns as
$p_{row}(i)\equiv p(F_{i,1})$ and
$p_{col}(j)\equiv p(F_{1,j})$, respectively, where
 $p(F_{i,j})$ is the Grassmann parity of the matrix element $F_{i,j}$.
For the matrices $L_{2m}^\pm$ one has $p_{row}(i)=p_{col}(i)\equiv p(i)$.
Matrix $F$ has certain Grassmann parity, if the expression
\ba\label{F-par}
p(F)=p(i)+p(j)+p(F_{i,j})
\ea
does not depend on $i$ and $j$. For even $n=2m$ the matrices $L_{n}^\pm$
have Grassmann parity $p(L_{2m}^\pm)=0$, while for odd $n=2m+1$ matrices
$L_{2m+1}^\pm$ have no definite parity and, therefore, for odd $n$
the zero-curvature
representation  (\ref{Ln-eqn}) (as well as the Lax pair
representation in one-dimensional space) does not make sense.

\subsection{Bi-Hamiltonian structure of the periodic 1D generalized
fermionic Toda lattice hierarchy}

The periodic 1D generalized fermionic Toda lattice equations (\ref{TODA-1d})
with the $2m$-periodic boundary
conditions $IV)$ (\ref{TODA-bound}) can be reproduced via the following Lax pair
representation:
\ba\label{Lax-eq-2m}
\partial L_{2m}=[L_{2m},L^-_{2m}]\, , \ \ \ \ \ L_{2m}=L^+_{2m}+L^-_{2m},
\ea
\vspace*{-.8cm}
\ba
 L_{2m}&=&\left(
\begin{array}{cccccccccc}
c_1&\gamma_1&1&0&\dots&\dots&0&0&d_1/w&\rho_1/w\\
\rho_2&c_2&\gamma_2&1&\dots&\dots&0&0&0&d_2/w\\
d_3&\rho_3&c_3&\gamma_3&\dots&\dots&0&0&0&0\\
0&d_4&\rho_4&c_4&\dots&\dots&0&0&0&0\\
&&&&\dots&\dots&\dots&&&\\
&&&&\dots&\dots&\dots&&&\\
0&0&0&0&\dots&\dots&c_{2m-3}&\gamma_{2m-3}&1&0\\
0&0&0&0&\dots&\dots&\rho_{2m-2}&c_{2m-2}&\gamma_{2m-2}&1\\
w&0&0&0&\dots&\dots&d_{2m-1}&\rho_{2m-1}&c_{2m-1}&\gamma_{2m-1}\\
w \gamma_{2m}&w&0&0&\dots&\dots&0&d_{2m}&\rho_{2m}&c_{2m}\
\end{array}
\right). \nn
\ea
where $L_{2m}^\pm$ are defined according  to eqs. (\ref{Ln-pair}).

The $2m$-periodic 1D generalized fermionic Toda lattice equations
(\ref{TODA-1d}) possess the bi-Hamiltonian structure which can
easily be derived
from the first and second Hamiltonian structures (\ref{1pb-TODA}) and
(\ref{2pb-TODA}), if one makes changes there
according to the substitution
\ba\label{per-rule}
\delta_{i,j+k}\to \delta_{i,j+k}+\delta_{i,j-2m+k},\ \ \ \ \ \
\delta_{i,j- k}\to \delta_{i,j- k}+\delta_{i,j+2m-k}, \ \ \ \ \ (k>0)
\ea
and change the sum limits in the Hamiltonians (\ref{HAM-bos}) as
\ba\label{Ham-2m}
 &&H_1^{2m} = \sum_{i =1 }^{2m} (-1)^i c_i,\ \ \ \ \ \
H_2^{2m} = \sum_{i =1}^{2m} (-1)^i (  \frac12 c_i^2 +d_i+
 \rho_i \gamma_{i- 1}).
\ea
Thus, the first and second Hamiltonians structures explicitly are
 \ba
\{d_i, c_j\}_1& =&  ( -1 ) ^j d_i( \delta_{i, j+2}  - \delta_{i, j }+
\delta_{i, j-2m +2} ),\nn \\
\{c_i,\rho_j\}_1& =&  ( -1 ) ^j \rho_j
        (\delta_{i, j - 1} +  \delta_{i, j}+\delta_{i, j+2m -1} ),\nn \\
  \{\rho_i,
      \rho_j\}_1 &=&      ( -1 ) ^j (d_i (\delta_{i,
              j + 1}+\delta_{i, j-2m +1} )\
- d_j (\delta_{i,j - 1}+\delta_{i, j+2m -1} ))   ,\nn\\
  \{\gamma_i, \gamma_j\}_1& =&
     ( - 1 ) ^j ( \delta_{i, j + 1}- \delta_{i,j - 1}+
\delta_{i, j-2m +1}-\delta_{i, j+2m -1}  )
\ea
 and
\ba
  \{d_i,d_j\}_2&=&(-1)^j d_i d_j(\delta_{i,j+2}-\delta_{i,j-2}+
\delta_{i,j-2m+2}-\delta_{i,j+2m-2}),\nn \\
  \{d_i, c_j\}_2&=&( -1 )^j d_i c_j ( \delta_{i, j+2}- \delta_{i,j}+
\delta_{i,j-2m+2}),\nn \\
  \{c_i,c_j\}_2&=&(-1)^j (d_i (\delta_{i, j + 2} +\delta_{i,j-2m+2})
-d_j (\delta_{i, j - 2} +\delta_{i,j+2m-2})\nn\\
&&-
            \gamma_j  \rho_i (\delta_{i, j+1}+\delta_{i,j-2m+1})-
                 \gamma_i \rho_j (\delta_{i, j - 1}+\delta_{i,j+2m-1})),\nn\\
 \{d_i,\rho_j\}_2&=&(-1)^j  d_i \rho_j (\delta_{i,j+2}+\delta_{i,j-1}+
\delta_{i,j-2m+2}+\delta_{i,j+2m-1}),\nn\\
  \{d_i,\gamma_j\}_2&=&(-1)^j d_i
        \gamma_j  ( \delta_{i, j + 2} +\delta_{i, j + 1}+
\delta_{i,j-2m+2}+\delta_{i,j-2m+1}),\nn\\
  \{c_i,\rho_j\}_2&=&(-1)^j (c_i  \rho_j (\delta_{i,j}+\delta_{i,j-1}+
\delta_{i,j+2m-1})\nn\\
&&-
     d_j  \gamma_i (\delta_{i, j - 2}+ \delta_{i,j+2m-2})
        - d_i  \gamma_j\ (\delta_{i, j + 1}+\delta_{i,j-2m+1})),\nn\\
  \{c_i,\gamma_j\}_2&=&(-1)^j (\rho_i \delta_{i, j + 2} + \rho_j
        \delta_{i, j - 1}+\delta_{i,j-2m+2}+\delta_{i,j+2m-1}),\nn\\
 \{\rho_i,\gamma_j\}_2&=&(-1)^j  (\rho_i  \gamma_j (\delta_{i, j + 1}+
        \delta_{i,j-2m+1})+
              d_i (\delta_{i,j + 3}+\delta_{i,j-2m+3}) -
      d_j (\delta_{i, j - 1}+\delta_{i,j+2m-1}),\nn\\
\{\rho_i,
      \rho_j\}_2&=&(-1)^j ((\rho_i  \rho_j - d_j  c_i)(\delta_{i,j - 1}+
      \delta_{i,j+2m-1} )+
     (\rho_i  \rho_j + d_i  c_j) \ (\delta_{i, j + 1}+
      \delta_{i,j-2m+1})),\nn\\
 \{\gamma_i,\gamma_j\}_2&=&( -1 ) ^j(  c_i (
          \delta_{i, j+1}+\delta_{i,j-2m+1})- c_j(\delta_{i,j - 1}+
      \delta_{i,j+2m-1})),
\ea
respectively.

Bosonic integrals of motion
of the $2m$-periodic 1D generalized Toda lattice hierarchy can be
derived via the following general formula:
\begin{equation}\label{str-per}
 str L^k_{2m}= \sum_{p=1}^{2m} (-1)^p (L_{2m}^k)_{pp} =k (H_k^{2m}+
w I_{k-m}^{2m}+\widehat I_{k+m}^{2m}/w
+\delta_{k,2m} \widehat I_{4m}^{2m}/w^2
),  \ \ \ \ \ \ k=1\dots 2m.
\end{equation}
Here $H_k^{2m}$ are the bosonic Hamiltonians (\ref{Ham-2m}),
and $I_k^{2m}$ and $\widehat I_k^{2m}$ are the additional conserved  quantities.
We analyzed attentively the quantities $I_k^{2m}$ for the case $m=2,3$ and
found that $I_k^{2m}$ can be decomposed into a sum of a few terms which
are conserved separately and besides $H^{2m}_k$ contain two more independent
integrals of motion of length dimension $k<-1$
\ba \label{coef-str}
I^{2m}_p&=&0,\mbox{  if   } p\le 0,\nn\\
I_1^{2m}&=&H_1^{2m}+1/2 \ S_{3}^{2m}S_{4}^{2m},\nn\\
I_2^{2m}&=&H_2^{2m}+U_2^{2m}+V_2^{2m}+1/2 \ H_1^{2m} S_{3}^{2m}S_{4}^{2m},
\nn\\
\widehat I_{m+1}^{2m}&=&0, \ \ \ \
\widehat I_{2m}^{2m}=U_{2m}^{2m}-V_{2m}^{2m}
\ea
where
$U_k^{2m}$ and $V_k^{2m}$ are additional bosonic integrals of motion
\ba
U_2^{2m}&=&\sum\limits_{j=1}^{m}\sum\limits_{i=1}^{m}\sum\limits_{k=1}^{m}
c_{2j+1} \gamma_{2i}\gamma_{2k-1}+
\frac14 \sum\limits_{j=1}^{2m}\sum\limits_{i=0}^{m-2}
\sum\limits_{k=i+1}^{m-1}\sum\limits_{p=k}^{m-1}
(-1)^j \gamma_j \gamma_{j+2i+1}\gamma_{j+2k}\gamma_{j+2p+1},\nn\\
V_2^{2m}&=&\sum\limits_{j=1}^{2m}
\sum\limits_{i=1}^{m} (-1)^j \Bigl (
 \frac12 c_j c_{j+2i}-\gamma_{j-1}\rho_{j+2i}\Bigr )\nn\\
&&+\sum\limits_{j=1}^{m}
\Bigl( c_{2j} \sum\limits_{i=0}^{m-2} \sum\limits_{k=i}^{m-2}
\gamma_{2j+2i+1} \gamma_{2j+2k+2}-
 c_{2j-1} \sum\limits_{i=0}^{m-1} \sum\limits_{k=i}^{m-1}
\gamma_{2j+2i-1} \gamma_{2j+2k} \Bigr),\nn\\
U_{2m}^{2m}&=&\prod\limits_{i=1}^m d_{2i},\ \ \ \ \ \ \    \
V_{2m}^{2m}\ =\ \prod\limits_{i=1}^m d_{2i-1}\ \ \ \ \ \ \    \
\label{fff}
\ea
and $S_{3}^{2m}$ and $S_{4}^{2m}$ are fermionic integrals (see eqs. \p{Ham-fer-2m}).
Our conjecture is that formulae \p{coef-str}--\p{fff} are valid not
only for the values $m=2,3$ for which they were actually calculated,
but also for an arbitrary value of $m$.

   The first fermionic Hamiltonians \p{HAM-fer}
in the $2m$-periodic case become
\ba\label{Ham-fer-2m}
S_{1}^{2m} = \sum_{i = 1}^{2m}(-1)^i \rho_i g_i^{-1},\ \ \ \ \
  S_{2}^{2m} = \sum_{i = 1} ^{2m}
  \rho_i g_i^{-1},\ \ \ \ \ \ \
  S_{3}^{2m} = \sum_{i=1}^{2m} (-1)^i \gamma_i ,\ \ \ \ \ \ \
S_{4}^{2m} = \sum_{ i = 1} ^{2m}  \gamma_i.\ \
\ea
Note that in the periodic case the fields $g_j$ are connected with the
fields $d_j$ via the irreversible relation $d_j=g_jg_{j-1}$.  For the fields
$g_j$ there are equations (\ref{g-eq})
and it seems reasonable to consider (\ref{TODA-1d}),
(\ref{TODA-1d-susy}) and (\ref{g-eq}) as a single joined system of equations.
In this case, the system possesses the $N=4$ supersymmetry and has
additional bosonic
integrals of motion  $\widehat V_k^{2m}$ and $\widehat U_k^{2m}$
which can be derived using automorphism (\ref{toda-auto})
\ba
\widehat V_k^{2m}&=&V_k^{2m}(\gamma_j\to (-1)^j\rho_{j+1} g^{-1}_{j+1},
\rho_j\to (-1)^j\gamma_{j-1} g_j),\ \ \ \ \ \nn\\
\widehat U_k^{2m}&=&U_k^{2m}(\gamma_j\to (-1)^j\rho_{j+1} g^{-1}_{j+1},
\rho_j\to (-1)^j\gamma_{j-1} g_j).
\ea

We suppose that
the Hamiltonians (\ref{Ham-fer-2m}) are the only independent fermionic
integrals of motion which exist for the $2m$-periodic 1D generalized
fermionic Toda lattice equations (\ref{TODA-1d}).
Thus, we have checked that higher fermionic Hamiltonians
of length dimensions -3/2 and -5/2
in the $2m$-periodic
case for $m=2$ become composite and can be expressed via the fermionic
Hamiltonians (\ref{Ham-fer-2m}) and bosonic integrals of motion
as a sum of composite terms.

\subsection{The r-matrix formalism}

There is another approach to reproduce bosonic integrals of motion \cite{FT}.
Let us consider the $2m$-periodic auxiliary linear problem
\ba\label{lin-pr}
\lambda\psi_j=(L_{2m})_{ij}\psi_j\equiv
\rho_j\psi_{j-1}+d_j\psi_{j-2}+\psi_{j+2}+
\gamma_j\psi_{j+1}+c_j\psi_j
\ea
\ba\label{lin-pr-time}
\partial \psi_j=(L^{-}_{2m})_{ij}\psi_j\equiv
\rho_j\psi_{j-1}+d_j\psi_{j-2}
\ea
for the wave functions $\psi_j$ such
that $\psi_{j+2m}=w\psi_j$.
One can check that (\ref{lin-pr})--\p{lin-pr-time} are equivalent to the
following linear problem:
\ba \label{lin-pr-FF}
 \Phi_{j+1}={\mathfrak L}_j(\lambda) \Phi_{j}, \quad
 \partial \Phi_{j}={\mathfrak U}_j(\lambda) \Phi_{j},
\ \ \ \ \ \
\Phi_j=\left(
\begin{array}{l}
\psi_{j+1}\\
\psi_{j-1}\\
\psi_{j}\\
\psi_{j-2}
\end{array}
\right)
\ea
where
\ba
\label{laxlll}
{\mathfrak L}_j(\lambda)=\left(
\begin{array}{cccc}
-\gamma_j &-\rho_j&\lambda-c_j &-d_j\\
0&0&1&0\\ 1&0&0&0\\
0&1&0&0
\end{array}
\right),
 \ \
{\mathfrak U}_j(\lambda)=\left(
\begin{array}{cccc}
0 &-d_{j+1}&-\rho_{j+1}& 0\\
1&c_{j-1}-\lambda&\gamma_{j-1}&0\\
0&-\rho_j&0&-d_j\\
0&\gamma_{j-2}&1&c_{j-2}-\lambda
\end{array}
\right)
\ea
and the 1D generalized fermionic Toda lattice equations (\ref{TODA-1d})
result from the consistency condition
\ba\label{cons-FF}
\partial {\mathfrak L}_j(\lambda)=
{\mathfrak U}_{j+1}(\lambda){\mathfrak L}_j(\lambda)-
{\mathfrak L}_j(\lambda){\mathfrak U_j}(\lambda)
\ea
of the linear system (\ref{lin-pr-FF}).
Let us note that the $4\times 4$-matrix Lax operator
${\mathfrak L}_j(\lambda)$ \p{laxlll} has the
fermionic Grassmann
parity $p({\mathfrak L}_j(\lambda))=1$, according to the definition
(\ref{F-par}).
The transformations (\ref{ab-basis}) to the new basis
$\{a_j,\bar a_j,b_j,\bar b_j,\alpha_j,\bar \alpha_j,\beta_j,\bar \beta_j\}$
in the space of the functions $\{c_j,d_j,\rho_j,\gamma_j\}$
together with the new definitions
\ba\label{doubleLax}
\begin{array}{lc}
{\cal L}_j(\lambda)\equiv
{\mathfrak L}_{2j+1}(\lambda){\mathfrak L}_{2j}(\lambda), \quad
V_j\equiv {\mathfrak U}_{2j}(\lambda), \ \ \
&
F_j=\left (
\begin{array}{l}
\phi_{j}\\
\phi_{j-1}\\
\varphi_{j}\\
\varphi_{j-1}
\end{array}
\right)
\ \equiv \
 \Phi_{2j}    \ = \
\left(
\begin{array}{l}
\psi_{2j+1}\\
\psi_{2j-1} \\
\psi_{2j}\\
\psi_{2j-2}
\end{array}
\right )
\end{array}
\ea
 allow us to rewrite eqs. (\ref{lin-pr})--(\ref{cons-FF})
in the following equivalent form:
\ba\label{lin-pr-F}
&&\bar\beta_j\phi_{j-1}+\bar b_j
\varphi_{j-1}+\varphi_{j+1}-\bar\alpha_j\phi_j+ (\bar
a_j-\lambda)\varphi_j=0,\nn\\
&&\beta_j\varphi_{j}+b_j
\phi_{j-1}+\phi_{j+1}+\alpha_{j+1}\varphi_{j+1}+ (
a_j-\lambda)\phi_j=0,
\ea
\ba\label{all-pr-F}
 F_{j+1}={\cal L}_j(\lambda) F_{j},\ \ \ \
\partial F_{j+1}={V}_j(\lambda) F_{j+1}, \ \ \
\partial {\cal L}_j(\lambda)={ V}_{j+1}(\lambda){\cal L}_j(\lambda)-
{\cal L}_j(\lambda){ V_j}(\lambda)
\ea
where
\ba
{\cal L}_j(\lambda)=\left(
\begin{array}{cccc}
\lambda-\alpha_{j+1}\bar\alpha_j-a_j&\alpha_{j+1}\bar\beta_j-b_j&
(\bar a_j-\lambda)\alpha_{j+1}-\beta_j&
\bar b_j -\alpha_{j+1}\\
1&0&0&0\\
\bar\alpha_j&
-\bar\beta_j&\lambda-\bar a_j &-\bar b_j\\
0&0&1&0
\end{array}
\right), \\
 \ \         V_j(\lambda)=         \left(
\begin{array}{cccc}
0&-b_j&-\beta_j&0\\
1&a_{j-1}-\lambda &\alpha_j&0\\
0&-\bar\beta_j&0&-\bar b_j\\
0&-\bar\alpha_{j-1}&1&\bar a_{j-1}-\lambda
\end{array}
\right).
~~~~~~~~~~~~~~~~~~~~~~~~~~~~
\ea
Now, we introduce a new basis
$\{p_j,\bar p_j,x_j,\bar x_j,\eta_j,\bar \eta_j,\xi_j,\bar \xi_j\}$
in the space of the functions\\
$\{a_j$$,\bar a_j$$,b_j$$,\bar b_j$$,\alpha_j,\bar \alpha_j,\beta_j,\bar \beta_j\}$,
\begin{eqnarray}\label{ab-basis-can}
&&  a_i = p_i,\ \ \ \
  b_i= e^{x_i - x_{i - 1}},\ \ \ \
  \alpha_i = \eta_{i - 1} - \eta_i,\ \ \ \
  \beta_i= e^{x_i-\bar x_i}  \xi_i,\nn \\
&& \bar a_i =- \bar p_i,\ \ \ \
  \bar b_i= e^{\bar x_i - \bar x_{i - 1}},\ \ \ \
  \bar\alpha_i=-\bar\eta_i,\ \ \ \
  \bar\beta_i= e^{-x_{i - 1}+\bar x_i}(\bar\xi_i-\bar \xi_{j-1}),
\end{eqnarray}
such that the first Hamiltonian structure  (\ref{1pb-TODA})
becomes canonical
\ba\label{ab-pb-can}
\{x_i,p_j\}_1=\delta_{i,j},\ \ \
\{\bar x_i,\bar p_j\}_1=\delta_{i,j},\ \ \
\{\xi_i,\bar \xi_j\}_1=\delta_{i,j},\ \ \
\{\eta_i,\bar \eta_j\}_1=\delta_{i,j}
\ea
and after gauge transformation
\ba\label{omega-gauge}
F_j=\Omega_j \widetilde F_j, \quad
\Omega_j=\left(
\begin{array}{cccc}
1&0&\eta_j&0\\
0&-e^{x_{j-1}}&0&0\\
0&0&1&0\\
0&e^{\bar x_{j-1}}\bar\xi_{j-1}&0&-e^{\bar x_{j-1}}
\end{array}
\right)
\ea
the linear problem in eqs.
(\ref{all-pr-F}) looks like
$$ \widetilde F_{j+1}=\widetilde {\cal L}_j(\lambda)\widetilde F_j$$
 where all matrix entries of
the matrix $\widetilde {\cal L}_j(\lambda)$ are
defined at {\it the same lattice node(!)}
\ba\label{L-tilde}
\widetilde {\cal L}_j(\lambda)\! =\!
\Omega^{-1}_{j+1}{\cal L}_j(\lambda)\Omega_j\! =\!
\left(\! \! \! \!
\begin{array}{cccc}
 \lambda- p_j +\eta_j\bar\eta_j&e^{x_j}+e^{\bar x_j}\bar\xi_j\eta_j&
- e^{x_j-\bar x_j}\xi_j -(p_j+\bar p_j)\eta_j&-e^{\bar x_j}\eta_j\\
-e^{-x_j}&0&-e^{-x_j}\eta_j&0\\
-\bar\eta_j&  e^{\bar x_j}\bar\xi_j
& \lambda+\bar p_j+\eta_j\bar\eta_j &e^{\bar x_j}\\
 e^{- x_j}\bar\xi_j&0&
-e^{-\bar x_j}+e^{-x_j}\bar\xi_j\eta_j&0
\end{array}
\! \! \! \!\right).~
\ea
Here, we note that the
$(4\times 4)$-matrix Lax operator
$\widetilde {\cal L}_j(\lambda)$ \p{L-tilde} has the bosonic Grassmann
parity $p(\widetilde {\cal L}_j(\lambda))=0$, according to the definition
(\ref{F-par}), and
\ba\label{sdet-L-cal}
 sdet \widetilde {\cal L}_j(\lambda)=1.
\ea

The matrices $\widetilde {\cal L}_j(\lambda)$ obey the $r$-matrix
Poisson brackets which are equivalent to the algebra (\ref{ab-pb-can})
\ba\label{rPB-L-tilde}
\{\widetilde {\cal L}_i(\lambda)\stackrel{\otimes}{,}\widetilde {\cal L}_j(\mu)\}=
[r(\lambda-\mu),\widetilde {\cal L}_i(\lambda)
\stackrel{\otimes}{\phantom{,}}
\widetilde {\cal L}_i(\mu)]\delta_{i,j}
\ea
where
\ba
r(\lambda-\mu)=\frac{P}{\mu-\lambda}
\ea
and
$$P_{ij;kl}=(-1)^{p(i)p(j)} \delta_{i,l}\delta_{j,k}$$
is the permutation matrix. The Grassmann parity function
 $p(j)=0(1)$ for bosonic (fermionic) rows and columns
 of a supermatrix, and
for the supermatrix $\widetilde {\cal L}_j (\lambda)$ (\ref{L-tilde})
we have $p(1)=p(2)=0$, $p(3)=p(4)=1$. In \p{rPB-L-tilde}
we have used the graded tensor product of two even supermatrices
$A$ and $B$ \cite{KulSk}
\ba(A
\stackrel{\otimes}{\phantom{,}}B)_{ij;kl}=(-1)^{p(j)(p(i)+p(k))}A_{ik}B_{jl}
\ea
with the properties
\ba\label{dir-prod-prop}
A\stackrel{\otimes}{\phantom{,}} B&=& P\, (B\stackrel{\otimes}{\phantom{,}} A)\, P,\nn\\
\{A\stackrel{\otimes}{,} B\}&=&- P\, \{B\stackrel{\otimes}{,} A\}\, P ,\nn\\
\{A\stackrel{\otimes}{,} BC\}&=&
 \{A\stackrel{\otimes}{,} B\}(I\stackrel{\otimes}{\phantom{,}} C)+
(I\stackrel{\otimes}{\phantom{,}} B) \{A\stackrel{\otimes}{,} C\}.
\ea

As a consequence of (\ref{rPB-L-tilde}) and (\ref{dir-prod-prop})
 the monodromy matrix
\ba\label{T-tilde}
\widetilde T_m(\lambda)&=&\prod_{j=1}^{\stackrel{\curvearrowleft}{m}}\widetilde
{\cal L}_j(\lambda)
\ea
satisfies the following Poisson bracket relation:
\ba\label{rPB-T-tilde}
\{\widetilde T_m(\lambda)\stackrel{\otimes}{,}\widetilde T_m(\mu)\}&=&
[r(\lambda-\mu),\widetilde T_m(\lambda)
\stackrel{\otimes}{\phantom{,}}\widetilde T_m(\mu)].
\ea
It follows from (\ref{rPB-T-tilde}) that $m$
 bosonic integrals of motion are in involution since
$$str \widetilde T_m(\lambda)=
(\widetilde T_m(\lambda))_{11}
+(\widetilde T_m(\lambda))_{22}
-(\widetilde T_m(\lambda))_{33}
-(\widetilde T_m(\lambda))_{44}
$$
is a polynomial of degree $m$ in $\lambda$ with integrals of motion as
the coefficient-functions and
\ba\label{tr-T-tilde}
 \{str \widetilde T_m(\lambda),str\widetilde T_m(\mu)\}=
str\{ \widetilde T_m(\lambda)\stackrel{\otimes}{,}\widetilde T_m(\mu)\}=
str[r(\lambda-\mu),\widetilde
T_m(\lambda)\stackrel{\otimes}{\phantom{,}}\widetilde T_m(\mu)]=0.
\ea

Let us note that the
operator ${\cal L}_j(\lambda)$ (\ref{doubleLax}) can be represented
as a product of two fermionic operators
${ l}_j(\lambda)$ and
$\bar{ l}_j(\lambda)$
\ba
{\cal L}_j(\lambda)=
{ l}_j(\lambda)
\bar{ l}_j(\lambda),\ \ \ \
{ l}_j(\lambda)=
{\mathfrak L}_{2j+1}(\lambda)W_j,\ \ \ \
\bar{ l}_j(\lambda)=W_j^{-1}
{\mathfrak L}_{2j}(\lambda),
\ea
where we have introduced the supermatrix $W_j$ which we define as
\ba
W_j=\left (
\begin{array}{cccc}
1&0&0&0\\
0&1&0&0\\
0&0&1&0\\
0&0&0&e^{x_j-1}
\end{array}
\right ).
\ea
Then, after the gauge transformation
(\ref{omega-gauge}) the Lax operator
$\widetilde {\cal L}_j(\lambda)$
(\ref{L-tilde}) has the form of the product of
two fermionic operators
$\widetilde{ l}_j(\lambda)$ and
$\widetilde{\bar{ l}}_j(\lambda)$
\ba
&&\quad \quad \quad \quad \quad \quad \quad \quad \quad
\quad \quad \quad \widetilde {\cal L}_j(\lambda) =
\widetilde{ l}_j(\lambda)
\widetilde{\bar{ l}}_j(\lambda),\ \ \ \ \nonumber\\
&&\widetilde{ l}_j(\lambda)=
\Omega^{-1}_{j+1}{ l}_j(\lambda)\equiv
\Omega^{-1}_{j+1}{\mathfrak L}_{2j+1}(\lambda)W_j, \ \ \ \
\widetilde{\bar{ l}}_j(\lambda)=\bar{ l}_j(\lambda)\Omega_{j}\equiv
W_j^{-1}{\mathfrak L}_{2j}(\lambda)\Omega_{j}
\ea
and each of them is defined at {\it the same lattice node(!)}
\ba
\widetilde{ l}_j(\lambda)&=&
\left(
\begin{array}{cccc}
-\eta_j&-e^{x_j-\bar x_j}\xi_j&\lambda-p_j&-e^{x_j}\\
0&0&-e^{-x_j}&0\\
1&0&0&0\\
0&-e^{\bar x_j}&e^{ x_j}\bar\xi_j&0
\end{array}
\right ),\nn\\
\widetilde{\bar{ l}}_j(\lambda)&=&
\left(
\begin{array}{cccc}
-\bar\eta_j&e^{\bar x_j}\bar\xi_j&\lambda+\bar p_j+\eta_j\bar\eta_j
&e^{\bar x_j}\\
0&0&1&0\\
1&0&\eta_j&0\\
0&-1&0&0
\end{array}
\right ).
\ea
It would be interesting to establish $r$-matrix Poisson bracket relations
(if any) between the fermionic supermatrices
$\widetilde{ l}_j(\lambda)$,
$\widetilde{\bar{ l}}_j(\lambda)$.

In order to rewrite the monodromy matrix in terms of the original fields
$\{d_j,c_j,\rho_j,\gamma_j\}$ one can perform
  the inverse
gauge transformations
$${\cal L}_j(\lambda)=\Omega_{j+1}
\widetilde {\cal L}_j(\lambda)\Omega^{-1}_j$$
and define the monodromy matrix
\ba\label{mon-mat}
T_m(\lambda)&=&
\prod_{j=1}^{\stackrel{\curvearrowleft}{m}}
 {\cal L}_j(\lambda)=
\Omega_1\left(\prod_{j=1}^{\stackrel{\curvearrowleft}{m}}
\widetilde {\cal L}_j(\lambda)\right ) \Omega_1^{-1}=
\Omega_1
{\widetilde T}_m(\lambda) \Omega_1^{-1}
\ea
where all the matrix entries are expressed in terms of the fields
$\{d_j,c_j,\rho_j,\gamma_j\}$.
In (\ref{mon-mat}) the periodicity property
$\Omega_{m+1}=\Omega_1$ of the gauge
transformation matrix (\ref{omega-gauge}) has been used.
Relation (\ref{tr-T-tilde})
is also true for the monodromy matrix $T(\lambda)$
because of the relation
 $str T_m^k(\lambda)=str\widetilde T_m^k(\lambda)$.
In other words, we have
shown that $m$ integrals of the motion being expressed in terms of the
original fields
$\{d_j,c_j,\rho_j,\gamma_j\}$ are in involution.
However, as the decomposition (\ref{str-per})
shows, for the $2m$-periodic problem there are more than $m$ integrals
of motion. In order to obtain them, let us investigate
the decomposition
\ba\label{T2-dec}
str T^2_m{(\lambda)}=\sum\limits_{k=0}^{2m-1} J_{2m-k}^{2m} \lambda^{k}.
\ea
The first several coefficients $J^{2m}_p$ have the following
explicit form:
\ba
J^{2m}_1&=& 2 H_1^{2m}-2\ S_3^{2m}S_4^{2m},\nn\\
J^{2m}_2&=& 2 H_2^{2m}-4 V_2^{2m}+8U_2^{2m}+H_1^{2m} S_3^{2m}S_4^{2m}.
\ea
One can see that the additional integrals $V_2^{2m}$ and  $U_2^{2m}$
are contained in the coefficient $J_2^{2m}$ in a different combination
than in $I_2^{2m}$ (\ref{coef-str}). This is the way to detect them.
We suppose that for integrals of higher length dimensions the
situation is the same: there are three independent coefficients
$H^{2m}_k$, $I^{2m}_k$  and $J^{2m}_k$
of length dimension $k$
in decompositions (\ref{str-per}) and (\ref{T2-dec}) and each of them
is an independent integral of motion.

Having bosonic and fermionic integrals of motion for the $2m$-periodic
1D generalized fermionic
Toda lattice equations (\ref{TODA-1d}) it is easy to obtain
integrals of motion for the periodic $N=4$ (\ref{g-toda-1d-1order}) and
$N=2$ (\ref{ab-toda-red}) Toda lattice equations.
This can be done, respectively,
using transformations (\ref{g-basis}) and (\ref{ab-basis}) together with the
reduction constraints \p{reds1} and \p{ab-sol}.

\subsection{Spectral curves}

The Lax operator $L_{2m}$ (\ref{Lax-eq-2m}) and monodromy matrix $T_m$
(\ref{mon-mat}) have the common spectrum
\ba
L_{2m}\psi=\lambda\psi, \ \ \ \ T_m\psi=w\psi;
\ea
so there exist relations ${h}(\lambda,w)=0$, ${h^{-1}}(\lambda,w)=0$
between them which are
formulated in terms of the characteristic function
\ba\label{T-det}
{h}(\lambda,w)=sdet(w-T_m(\lambda)).
\ea

Calculating the superdeterminant and applying the modified Euclidean
algorithm \cite{KN}
to an arbitrary supermatrix $M$ with the parities
$p(1)=
p(2)=0$,
$p(3)=
p(4)=1$
one can find
\ba\label{sp-curv-22}
sdet(w-M)
=\frac{w^2+\sigma_1 w+\sigma_2}
{w^2+\sigma_3 w+\sigma_4}
\ea
where all the coefficients $\sigma_m$ are expressed in
terms of  four invariants
$$\mu_k=str M^k=(M^k)_{11}+(M^k)_{22}-(M^k)_{33}-(M^k)_{44},
\quad k=1,2,3,4$$
of the  matrix $M$
\ba
\sigma_1=\sigma_3+\epsilon_1,\ \ \
\sigma_3=\frac{1}{\epsilon_1}(
\frac{\epsilon_4}{\epsilon_3 }+\epsilon_2),\ \ \
\sigma_2=\sigma_4+
\frac{\epsilon_4}{\epsilon_3 },\ \ \
\sigma_4=\frac{1}{\epsilon_1^2}(
\frac{\epsilon_2 \epsilon_4}{\epsilon_3 }+\epsilon_3),\nn
\ea
\vspace*{-.9cm}
\ba
&&\epsilon_1=-\mu_1,\ \ \
\epsilon_2=-1/2\ (\mu_1^2-\mu_2),\ \ \
\epsilon_3=1/12\ \mu_1^4+1/4\ \mu_2^2-1/3\ \mu_1\ \mu_3,\nn\\
&&\epsilon_4=
-1/4\ \mu_4\ \mu^2_1+\mu_2\ (1/24\ \mu_1^4-1/8\ \mu_2^2+1/3\ \mu_3\ \mu_1).
\label{matrixM}
\ea

Now let us adapt formulae \p{sp-curv-22}--\p{matrixM} obtained for
an arbitrary matrix $M$ to the case when $M$ is the monodromy matrix
$T_m(\lambda)$ (\ref{mon-mat}).
For the monodromy matrix $T_m(\lambda)$ there is a relation
\ba\label{sdetT}
sdet\ T_m(\lambda)=1
\ea
which is a consequence of (\ref{sdet-L-cal}) and
imposes constraints on the coefficients $\sigma_k$. Taking
$h(\lambda,w)$ at
$w=0$ and using (\ref{sdetT}) one finds
$\sigma_4^{T_m}=\sigma_2^{T_m}$, $\epsilon_4^{T_m}=0$ and
\ba\label{sp-cur-T}
h(\lambda,w)
=\frac{w^2+\sigma_1^{T_m} w+\sigma_4^{T_m}}
{w^2+\sigma_3^{T_m} w+\sigma_4^{T_m}}
\ea
where $\sigma_k^{T_m}=\sigma_k(\epsilon_s\to\epsilon_s^{T_m})$,
$\epsilon_4^{T_m}=0$,
$\epsilon_s^{T_m}=\epsilon_s(\mu_k\to str T^k_m(\lambda))$, $s=1,2,3$.
From eq. (\ref{sp-cur-T}) one can obtain two spectral curves as zeros
of the numerator and denominator
\ba\label{num-den}
{\cal P}_{num}(\lambda,w)=w^2+\sigma_1^{T_m}
w+\sigma_4^{T_m}=0,\nn\\
{\cal P}_{den}(\lambda,w)=w^2+\sigma_3^{T_m}
w+\sigma_4^{T_m}=0.
\ea
The eigenvalues $w$ of the curve ${\cal P}_{num}(\lambda,w)=0$
correspond to the  eigenvectors with the even Grassmann parity $p(\psi)=0$,
while the eigenvalues of the
curve ${\cal P}_{den}(\lambda,w)=0$ correspond to the odd
eigenvectors $p(\psi)=1$.

\subsection{Reduction: r-matrix approach and spectral curves for the
periodic 1D N=2 Toda lattice hierarchy}
For completeness, in this section
 we give a short summary of the $r$-matrix formalism for the
periodic 1D $N=2$ supersymmetric Toda lattice equations
(\ref{ab-toda-red}).
In the case under consideration the auxiliary linear problem
(\ref{lin-pr-F}),
being reduced by constraints \p{reds1} and \p{ab-sol},
becomes
\ba\label{lin-pr-N2}
&&\bar\beta_j\phi_{j-1}+\varphi_{j+1}-\bar\alpha_j\phi_j- (
\frac{\beta_j\bar \beta_j}{b_j}
+\lambda)\varphi_j=0,\nn\\
&&\beta_j\varphi_{j}+b_j
\phi_{j-1}+\phi_{j+1}+\alpha_{j+1}\varphi_{j+1}+ (
a_j-\lambda)\phi_j=0,
\ea
and it is equivalent to the
following linear problem:
\ba \label{lin-pr-F-N2}
\widehat F_{j+1}=\widehat {\cal L}_j(\lambda) \widehat F_j
\ea
where
\ba
\widehat{\cal L}_j(\lambda)=\left(
\begin{array}{ccc}
 \lambda-\alpha_{j+1}\bar\alpha_j-a_j& \alpha_{j+1}\bar\beta_j-b_j&
-(\displaystyle{
\frac{\beta_j\bar\beta_j}{b_j}}+\lambda)\alpha_{j+1}-\beta_j \\
1&0&0\\
\bar\alpha_j&-\bar\beta_j&
\lambda+\displaystyle{\frac{\beta_j\bar \beta_j}{b_j}}
\end{array}
\right),
 \ \ \
\widehat F_j=\left(
\begin{array}{l}
\phi_j\\
\phi_{j-1}\\
\varphi_j
\end{array}
\right).~~
\ea
As concerns the periodic 1D $N=2$ Toda lattice equations
(\ref{ab-toda-red}), they are equivalent to the lattice
zero-curvature representation
\ba
\partial {{\widehat{\cal L}}}_j(\lambda)={ {\widehat V}}_{j+1}
(\lambda){ {\widehat{\cal
L}}}_j(\lambda)- {{\widehat {\cal L}}}_j
(\lambda){ {\widehat V}}_j(\lambda)
\ea
with
\ba
{ {\widehat V}}_j=\left(
\begin{array}{ccc}
0&-b_j&-\bar\beta_j\\
0&-\lambda-a_{j-1}&\alpha_j\\
0&-\bar\beta_j&0
\end{array}\right ).
\ea

In the canonical basis $\{x_j,p_j,\alpha_j,\bar\alpha_j,\beta_j,
\bar\beta_j\}$ (\ref{ab-can})
after the gauge transformation
\ba
\widehat { F_j}=\widehat \Omega_j\widehat {\widetilde F_j}, \quad
\widehat \Omega_j=\left( \begin{array}{ccc} 1&0&\eta_j\\
  0&-e^{ x_{j-1}}&0\\
0&0&1
\end{array}
\right)
\ea
 eqs. (\ref{lin-pr-F-N2}) take the form
 $$ \widehat {\widetilde F}_{j+1}=
 \widehat {\widetilde {\cal L}}_j (\lambda) \widehat {\widetilde F}_j $$
where
\ba\label{L-tilde-N2}
\widehat{\widetilde {\cal L}}_j(\lambda)=\widehat\Omega^{-1}_{j+1}
\widehat{{\cal L}}_j(\lambda)\widehat\Omega_j=
\left( \begin{array}{ccc}
\lambda+\eta_j\bar\eta_j-p_j&e^{x_j}+\bar\xi_j\eta_j&
 -e^{x_j}\xi_j -(p_j +\xi_j\bar\xi_j)\eta_j\\
-e^{-x_j}&0&
-e^{-x_j}\eta_j\\
-\bar\eta_j &\bar\xi_j&
\lambda+\xi_j\bar\xi_j +\eta_j\bar\eta_j
\end{array}
\right)
\ea
and is defined at the same lattice node.
The matrices $\widehat{\widetilde {\cal L}}_j(\lambda)$
have the Grassmann parities $p(1)=p(2)=0$, $p(3)=1$ and
obey the same $r$-matrix
Poisson bracket relations (\ref{rPB-L-tilde})
with the appropriate $r$ matrix.

The equation for eigenvalues
\ba
sdet(w-\widehat T_m(\lambda))=0 \quad or \quad \infty
\ea
of the monodromy matrix
\ba
{\widehat T}_m(\lambda)=
\widehat\Omega_1^{-1}
\widehat {\widetilde T}_m(\lambda)
\widehat\Omega_1
=
\widehat\Omega_1^{-1}
\left( \prod_{j=1}^{\stackrel{\curvearrowleft}{m}}\widehat{\widetilde
{\cal L}}_j(\lambda)\right )
\widehat\Omega_1
=
 \prod_{j=1}^{\stackrel{\curvearrowleft}{m}}\widehat
{\cal L}_j(\lambda),
\ea
is defined by the characteristic function
\ba
\widehat{h}(\lambda,w)=\frac{w^2+\widehat\sigma_1w
+\widehat\sigma_2}
{w +\widehat\sigma_3}
\ea
where all the coefficients are expressed in terms of the invariants
of the monodromy matrix
$$\widehat\mu_k=str \widehat T_m^k=(\widehat T_m^k)_{11}+
(\widehat T_m^k)_{22}-(\widehat T_m^k)_{33}, \quad k=1,2,3,$$
\ba
\widehat \sigma_1=\widehat \sigma_3+\widehat\epsilon_1,\ \ \
\widehat\sigma_2=\widehat\epsilon_1\widehat\sigma_3+\widehat\epsilon_2,\ \ \
\widehat\sigma_3=\widehat\epsilon_3/\widehat\epsilon_2,\nn
\ea
\vspace*{-.9cm}
\ba
&&\widehat \epsilon_1=-\widehat\mu_1,\ \ \
\widehat\epsilon_2=1/2\ (\widehat\mu_1^2-\widehat\mu_2),\ \ \
\widehat\epsilon_3=1/3\ \widehat \mu_3-1/2\ \widehat\mu_2 \widehat\mu_1+1/6\
\widehat\mu^3_1.
\ea
From the relation $sdet\ \widehat{\widetilde {\cal L}}_j(\lambda)=\lambda^{-1}$
it follows that
$sdet\ \widehat T_m=\lambda^{-m}$ and, consequently,
$\widehat\sigma_3=-\lambda^m\widehat\sigma_2$.

\subsection{Periodic Toda lattice equations in the canonical basis}
\label{perTL-can}
In the previous subsections we considered the 1D
generalized fermionic Toda lattice equations
(\ref{TODA-1d}) with the periodic boundary conditions $IV)$ (\ref{TODA-bound}).
All results obtained there can easily be transferred
to the case of the 1D $N=4$
(\ref{g-toda-1d-1order}) and $N=2$ (\ref{ab-toda-red})
Toda lattice equations after
transition to the new bases, (\ref{g-basis}) and (\ref{ab-basis}),
respectively, supplied with the reduction constraints \p{reds1}
and \p{ab-sol}. In particular, equations
(\ref{g-toda-1d-1order}) and (\ref{ab-toda-red}) with the periodic
boundary conditions are $N=4$ and $N=2$ supersymmetric, respectively,
and admit a bi-Hamiltonian representation which can be derived if one
changes the first and second Hamiltonian
structures (\ref{g-1PB})--(\ref{g-2PB}) and
(\ref{1pb-ab-basis-red})--(\ref{2pb-ab-basis-red}),
according to the rule (\ref{per-rule}).
In this subsection we consider equations
(\ref{g-toda-1d-1order}) and (\ref{ab-toda-red}) with the $2m$-periodic
and $m$-periodic
boundary conditions, respectively, in the canonical basis.

The  system (\ref{g-toda-1d-1order})
with the $2m$-periodic boundary conditions
in the canonical basis (\ref{g-basis-canon})
is quite similar to the infinite
system (\ref{g-toda-1d-1order}) in the canonical basis considered
in section \ref{N4-TL-can}.
Thus, the constraint
\ba
\prod_{k=1}^{2m} (-ig_k)=1
\ea
breaks the $N=4$ supersymmetry to the $N=2$ supersymmetry
and the $2m$-periodic Toda lattice equations (\ref{g-toda-1d-1order})
in the canonical basis have exactly the form (\ref{22-canon})
with the $N=2$ supersymetric flows (\ref{22-canon-susy}).
However, the $2m$-periodic $N=2$ Toda lattice equations (\ref{22-canon})
besides the $N=2$ supersymmetry
possess  additional four nonlocal fermionic nilpotent symmetries.
Let us present only nonzero flows which generate these
symmetries
\ba\label{22-nilp}
&&D_{s_1}x_j= \sum\limits_{k=1}^{2m}\left(
\chi_{k}^--(-1)^k\chi^+_{k}\right),\ \ \ \ \
D_{s_2}x_j= \sum\limits_{k=1}^{2m}\left(
\chi_{k}^++(-1)^k\chi^-_{k}\right),\ \ \ \ \
D_{s_3}\chi^+_j=
\sum\limits_{k=1}^{2m}p_k,
\nn\\
&&
D_{s_3}\chi^-_j=-(-1)^j
\sum\limits_{k=1}^{2m}p_k,\ \ \ \ \
D_{s_4}\chi^+_j= (-1)^j
\sum\limits_{k=1}^{2m}p_k,\ \ \ \ \
D_{s_4}\chi^-_j=
\sum\limits_{k=1}^{2m}p_k.
\ea
These flows
anticommute with each other and
with the sypersymmetric flows (\ref{22-canon-susy}) except
the following nonzero anticommutators:
\ba
\{\widetilde D_1,D_{s_2}\}= \partial_T, \ \ \ \
\{\widetilde D_1,D_{s_4}\}= -\partial_T, \ \ \ \
\{\widetilde D_2,D_{s_1}\}= \partial_T, \ \ \ \
\{\widetilde D_2,D_{s_3}\}= -\partial_T
\ea
where we have introduced the new evolution derivative
\ba
\partial_T q_j\equiv \{H_1^2,q_j\}_1
\ea
which gives nontrivial flows only for the fields $x_j$
\ba
\partial_{T}x_j=-2\sum\limits_{k=1}^{2m}p_k,\ \ \ \ \
\partial_{T}p_j=0. \ \ \ \ \
\partial_{T}\chi^{\pm}_j=0.
\ea

The $2m$-periodic $N=2$ Toda lattice equations (\ref{22-canon}) can be generated
using the Hamiltonian
\ba\label{}
H_2&=&\sum_{j=1}^{2m} (-1)^j (\frac12 p_j^2-e^{x_j-x_{j-2}}-
 e^{x_j-x_{j-1}} (\chi^-_{j-1}-(-1)^j \chi^+_j)(\chi^-_j+(-1)^j \chi^+_{j-1}))
\ea
and the canonical first Hamiltonian structure (\ref{22-1PB-canon}).
Following the standard procedure, one can derive the Lagrangian ${\cal L}$
and the action ${\cal S}$
\ba
{\cal S}&=&\int dt {\cal L}=\int dt [
\sum_{j=1}^{2m} p_j \frac{\partial}{\partial t}x_j+
\chi^-_j\frac{\partial}{\partial t}\chi_j^+-H_2]\nn\\
&=&\int dt \sum_{j=1}^{2m} [\frac12 (\frac{\partial}{\partial t}x_j)^2+
\chi^-_j\frac{\partial}{\partial t}\chi_j^+\nn\\
&&+
(-1)^j( e^{x_j-x_{j-2}}+
e^{x_j-x_{j-1}} (\chi^-_{j-1}-
(-1)^j \chi^+_j)(\chi^-_j+(-1)^j \chi^+_{j-1}))].
\ea
The variation of the action ${\cal S}$ with respect to the fields
$\{ x_j,\chi^-_j,\chi^+_j\}$ produces the equations of motion
(\ref{22-canon}) for them with reversed sign of time
($\partial \to -\frac{\partial}{\partial t}$)
where the momenta $p_j$
are replaced by $ (-1)^j \frac{\partial}{\partial t}x_j$.

The situation with the system (\ref{ab-toda-red}) with the $m$-periodic
boundary conditions is completely different. Let us recall that
the infinite system (\ref{ab-toda-red})
with the boundary
conditions $IIb)$ \p{ab-bc} for the fields $b_j$ at infinity
\ba\label{b-cond}
\lim_{j\to\pm\infty}b_j=1,
\ea
is not supersymmetric because the condition (\ref{b-cond}) spoils
the supersymmetric flows (\ref{ab-toda-red-susy}).
However, in the $m$-periodic case there is no condition (\ref{b-cond}),
and, as we will show, it is possible to build at least the $N=1$ supersymmetric
$2m$-periodic Toda lattice equations (\ref{ab-toda-red})
in the canonical basis.

The representation of the fields $b_j$ in the $m$-periodic canonical basis
(\ref{ab-can}) leads to the following constraint:
\ba
\prod_{k=j}^{m} b_j=1
\ea
and in order to preserve both supersymmetry flows
(\ref{ab-toda-red-susy}), one needs to  provide simultaneously two
additional constraints
\ba
\sum_{j=1}^m(\alpha_j\pm\bar\alpha_j)=0.
\ea
It appears that one can preserve
one supersymmetry if one modifies the transition to the canonical
basis (\ref{ab-can}) as follows:
\begin{eqnarray}\label{ab-s-canon}
&&  a_i = p_i,\ \ \ \
  b_i= e^{x_i - x_{i - 1}},\nn \\
&&  \beta_i= e^{x_i}  \xi_i, \ \ \ \
  \bar\beta_i= e^{-x_{i - 1}}\bar\xi_i,\nn \\
&&  \bar\alpha_i=-\bar\eta_i,\ \ \ \
  \alpha_i = \eta_{i - 1} - \eta_i +
        s(\bar\eta_{i -1}+\bar\eta_i),
\end{eqnarray}
where $s$ is an arbitrary parameter.
In this basis the $m$-periodic first Hamiltonian structure  still has
the canonical form (\ref{ab-1pb-can}), and using the Hamiltonian
\begin{eqnarray}
&&  H_2=- \sum_{i = 1}^n(\frac12 p_i^2  + e^{x_i - x_{i - 1}} +
          e^{-x_i}\bar\xi_{i + 1}(\eta_{i + 1} -
                \eta_i-s(\bar\eta_{i+1}+\bar\eta_i) )
 + e^{x_i}\xi_i \bar\eta_i
\end{eqnarray}
it generates the following equations:
\begin{eqnarray}\label{n-toda}
&&  \partial x_i= p_i, \ \ \
\partial\bar\xi_i= e^{x_i}\bar\eta_i, \ \ \
  \partial\xi_i=
   - e^{-x_{i - 1}}(\eta_{i - 1} -\eta_i +
          s(\bar\eta_{i}+\bar\eta_{i-1})) ,\nn \\
&&  \partial p_i= e^{x_{i + 1} - x_i} - e^{x_i - x_{i - 1}} -
      e^{x_i}\xi_i \bar\eta_i-
      e^{-x_i}
        \bar\xi_{i + 1} (\eta_i - \eta_{i + 1} +
           s(\bar\eta_i+\bar\eta_{i + 1})), \nn \\
&&\partial \eta_i=
   - e^{x_i} \xi_i +
      s( e^{-x_{i - 1}}\bar\xi_{i }+ e^{-x_{i }}\bar\xi_{i+1 }),  \ \ \
\partial\bar \eta_i= e^{-x_i}\bar\xi_{i + 1} -
      e^{-x_{i - 1}} \bar\xi_i.
\end{eqnarray}
One can standardly generate the Lagrangian ${\cal L}$
and the action ${\cal S}$
\ba
{\cal S}&=&\int dt {\cal L}=\int dt [
\sum_{j=1}^n p_j \frac{\partial}{\partial t}x_j+
\xi_j\frac{\partial}{\partial t}\bar\xi_j+
\eta_j\frac{\partial}{\partial t}\bar\eta_j
-H_2]\nn\\
&=&\int dt \sum_{j=1}^n
[-\frac12 (\frac{\partial}{\partial t}x_j)^2+
\xi_j\frac{\partial}{\partial t}\bar\xi_j+
\eta_j\frac{\partial}{\partial t}\bar\eta_j\nn\\
&&
+ e^{x_i - x_{i - 1}} +
          e^{-x_i}\bar\xi_{i + 1}(\eta_{i + 1} -
                \eta_i
-s (\bar\eta_{i + 1} +\bar  \eta_i )
) + e^{x_i}\xi_i \bar\eta_i ].
\ea
The variation of the action ${\cal S}$ with respect to the fields
$\{ x_j,\xi_j,\bar\xi_j,\eta_j,\bar\eta_j\}$ produces the equations of motion
(\ref{n-toda}) for them with reversed sign of time
($\partial \to -\frac{\partial}{\partial t}$)
where the momenta $p_j$
are replaced by $ - \frac{\partial}{\partial t}x_j$.
If, in addition to the momenta, the fields $\eta_j$ and $\bar \eta_j$
are also eliminated from (\ref{n-toda}) by means of the corresponding
equations expressing them in terms of the fields $\{x_j,\xi_j,\bar \xi_j\}$
and their derivatives, the remaining equations become
\ba\label{aa}
&&\partial^2x_j=e^{x_{j+1}-x_j}-e^{x_j-x_{j-1}}-\xi_j\partial \bar \xi_j+
\xi_{j+1}\partial \bar \xi_{j+1},\nn\\
&&\partial(e^{x_{j-1}}\partial\xi_j)=
e^{x_{j-1}}\xi_{j-1}-
e^{x_{j}}\xi_{j}, \ \ \ \ \
\partial(e^{-x_{j}}\partial\bar\xi_j)=
e^{-x_{j}}\bar\xi_{j+1}-
e^{-x_{j-1}}\bar\xi_{j}.
\ea
It is interesting to remark that the dependence of (\ref{aa})
on the parameter $s$ completely disappears.

For every $m$ the system (\ref{n-toda}) possesses the $N=1$
supersymmetry at the unique value $s=\pm 1/2$, and the supersymmetry
flows ($D_\pm^2=\mp\partial_t$) are
\begin{eqnarray}\label{xp-N1susy}
D_\pm x_i&=&-\eta_i  \pm 1/2
\sum_{k = 1}^{m - 1}\bar\eta_{i+k},\nn \\
D_\pm p_i&=& e^{x_i}\xi_i  \pm  e^{-x_i}\bar\xi_{i + 1},\nn \\
D_\pm\xi_i&=&\pm  e^{-x_{i - 1}}-\xi_i\eta_i \pm
\xi_i\bar\eta_i \pm 1/2 \sum_{k = 1}^{m - 1}
\xi_i\bar\eta_{i+k},\nn \\
D_\pm\bar\xi_i&=&e^{x_{i}}+\bar\xi_i\eta_i\mp\bar\xi_i\bar\eta_i\mp 1/2
\sum_{k = 1}^{m - 1}
\bar\xi_i\bar\eta_{i+k},\nn \\
D_\pm\eta_i&=& \pm p_i\pm
1/2 \sum_{k = 1}^{m - 1}
(p_{i + k}+\xi_{i + k }\bar\xi_{i + k }), \nn \\
D_\pm\bar\eta_i&=& p_i+\xi_i \bar\xi_i.
\end{eqnarray}

To close this section, let us only mention that
besides the supersymmetry flows \p{xp-N1susy} the system (\ref{n-toda})
possesses additional nilpotent symmetry
\begin{eqnarray}\label{xp-nilp}
D_p x_i&=&
\sum_{k = 1}^{m }\bar\eta_{k},\nn \\
D_p p_i&=&0,\nn \\
D_p\xi_i&=&\xi_i
\sum_{k = 1}^{m}\bar\eta_{k},\nn \\
D_p\bar\xi_i&=&-
\bar\xi_i\sum_{k = 1}^{m }  \bar\eta_{i},\nn \\
D_p\eta_i&=&
\sum_{k = 1}^{m }
(p_{ k }+\xi_{ k }\bar\xi_{ k }), \nn \\
D_p\bar\eta_i&=& 0.
\end{eqnarray}

{}~

{}~

\noindent{\bf Acknowledgments}
We would like to thank A.P. Isaev, P.P. Kulish, F. Magri and
M.A. Olshanetsky for useful discussions. We would also like to
thank P.P. Kulish for giving us reference
\cite{KN}. This work was partially supported by RFBR Grant No.
03-01-00781, RFBR-DFG Grant No. 02-02-04002,
DFG Grant 436 RUS 113/669 and by the Heisenberg-Landau program.

{}~

\end{document}